\documentclass[a4paper,12pt]{article}
\pdfoutput=1
\usepackage{jheppub}
\usepackage[toc,page]{appendix}
\usepackage{hyperref}
\usepackage{color}
\usepackage{graphicx,float}
\usepackage{ulem}
\usepackage{color}
\usepackage{overpic}
\usepackage{subcaption}

\usepackage[mathscr]{euscript}
\usepackage[usenames,dvipsnames]{xcolor}
\usepackage{amsmath,amsfonts,amssymb,verbatim,float,mathtools,bbold}
\usepackage{physics}
\usepackage{resizegather}
\usepackage{tikz}
\usetikzlibrary{shapes.misc}
 \tikzset{cross/.style={cross out, draw=black, minimum size=2*(#1-\pgflinewidth), inner sep=0pt, outer sep=0pt},
cross/.default={4pt}}

\usepackage{empheq}

\usepackage[most]{tcolorbox}
\tcbset{myformula/.style={colback=white, 
    colframe=black, 
    top=4pt,bottom=4pt,left=0pt,right=4pt,
    boxsep=0pt,
    arc=0pt,
    outer arc=0pt,
    fit algorithm=hybrid*
}}
\newcommand\fiteq[1]{%
  \sbox{\mybox}{$\displaystyle#1$}%
  \ifdim\wd\mybox>.85\textwidth\resizebox{.85\textwidth}{!}{\usebox{\mybox}}%
  \else\usebox{\mybox}\fi%
}
\newsavebox{\mybox}

\newtcolorbox{equationframe}{
math
}

\newcommand{\cov}{\nabla}
\newcommand{\p}{\partial}

\newcommand{\half}{{1\over 2}}
\newcommand{\OO}{\mathcal{O}}
\newcommand{\cA}{{\cal A}}

\newcommand{\la}{\langle}
\newcommand{\ra}{\rangle}

\newcommand{\pd}{\partial}

\def\half{{1 \over 2}}

\def\Or[#1]{{\text{O}}\left({#1}\right)}
\def\dotl[#1,#2]{\left\langle #1,\, #2 \right\rangle}
\def\dotlb[#1,#2]{\left\langle #1,\, #2 \right\rangle}
\def\dotlm[#1,#2]{\left[ #1,\, #2 \right]}
\def\dotp[#1,#2]{(\vect{#1} \cdot\vect{#2})}
\def\aff[#1,#2]{\hat{#1}(#2)}
\def\n4sym{{\cal N}=4 SYM}
\def\>{\rangle}
\def\<{\langle}
\def\({\left(}
\def\){\right)}
\def\weight[#1,#2,#3]{\{(#1),#2,#3\}}
\def\ads[#1]{$\text{AdS}_{#1}$}

\newcommand{\bp}{{\mathbf{p}}}

\newcommand{\be}{\begin{equation}}
\newcommand{\ee}{\end{equation}}
\newcommand{\beq}{\begin{eqnarray}}
\newcommand{\eeq}{\end{eqnarray}}
\newcommand{\ba}{\begin{align}}
\newcommand{\ea}{\end{align}}

\renewcommand{\vec}[1]{\boldsymbol{#1}}
\newcommand{\bs}{\begin{split}}
\def\sess\end{split}

\newcommand{\vect}[1]{{\boldsymbol{#1}}}

\definecolor{col1}{RGB}{153, 52, 121}

\makeatletter
\def\@fpheader{\relax}
\makeatother
\begin{document}

\title{Density response of holographic metallic IR fixed points with translational pseudo-spontaneous symmetry breaking}
\author{Aurelio Romero-Berm\'udez}
\affiliation{Instituut-Lorentz, $\Delta$ITP, Universiteit Leiden, P.O. Box 9506, 2300 RA Leiden, The Netherlands}

\emailAdd{romero@lorentz.leidenuniv.nl}
\abstract{
	The density response of charged liquids contains a collective excitation known as the plasmon.  In holographic systems with translational invariance the origin of this collective excitation is traced back to the presence of  zero-sound. Using a holographic model in which translational symmetry is broken pseudo-spontaneously, we show the density response is not dominated by a single isolated mode at low momentum and temperature.  As a consequence, the density response contains a broad asymmetric peak  with an attenuation which does not increase monotonically with momentum and temperature.
 }

\maketitle

\section{Introduction}

The density response of interacting charged liquids is expected to display a gapped propagating mode known as the plasmon. The plasmon may be understood by dressing a collective zero-sound excitation with the long-range electromagnetic interaction. In conventional metals (Fermi liquids), the plasmon is gapped and kinematically protected at low momentum because of the lack of available phase-space: the \textit{electron-hole} (Lindhard) continuum vanishes quadratically in momentum  \cite{Nozieres1999}. Therefore, in this low-energy regime the gapped plasmon is protected from decay. However, this paradigmatic picture  fails in strongly-correlated materials, like copper-oxides strange metals in the normal phase, across a wide range of doping. More specifically, optical  measurements suggest the plasmon is on the verge of being over-damped, i.e., the damping/attenuation rate is of the order of the plasmon frequency \cite{Bozovic1987,Slakey1991}. Similarly, over-damped dispersive plasmons were observed using energy transmission electron loss spectroscopy  \cite{Nucker1989,Nucker1991,Fink1994}. 
Recent experiments using low energy reflection electron loss spectroscopy with improved momentum resolution suggest the plasmon may be absent and the density response to be given by a featureless continuum \cite{Mitrano2017,Husain2019}. Clearly, this recent data is in tension with the optical measurements where the plasmon is still visible, albeit over-damped. Despite the controversy regarding the exact features of the plasmon in these materials, it is certain that they cannot be understood from the Fermi liquid wisdom. In this paper, we use  holography to study the density response in a locally quantum critical strange metal with infinite dynamical Lifshitz and hyperscaling violation exponents. In other words, the model has a flow to a metallic state in the IR with Sommerfeld entropy $s\sim T$. On the other hand, at high temperatures, the temperature scaling of the DC conductivity is typical of an electric insulator \cite{Amoretti2018}.  Moreover,  the model used here allows to study a phase in which translational symmetry is broken spontaneously, and by giving a small mass to the Goldstone, it also has a phase in which this symmetry is broken pseudo-spontaneously. However, it does not allow to study the transition between the two phases. (Pseudo)-spontaneous breaking patterns of translational symmetry have been studied recently in holography and field theory frameworks  \cite{Son2005,Nicolis2013,Baggioli2014,Andrade2017,Amoretti2017,Amoretti2017a,Amoretti2018,Alberte2018,Alberte2018a,Musso2018,Andrade2018,Donos2019}. In the holographic context, the focus has mainly been on the zero momentum optical conductivity. Here, we extend this analysis to finite momentum and focus on the density response, which may easily be related to the electric conductivity by a Ward identity.  Furthermore, in holography, the neutral density response and the holographic zero-sound excitation have also been extensively explored in various probe-brane models \cite{Karch:2009,Kulaxizi2008,Kulaxizi2009,Hoyos2010,Kaminski2010,Davison:2011}, back-reacted Dirac-Born-Infeld models \cite{Gushterov2018},  as well as in bottom-up effective models \cite{Edalati2010c,Davison2011b}. More recently, the effects of the boundary Coulomb interaction have also been considered in holographic setups to correctly account for plasmon physics \cite{Aronsson2017,Aronsson2018,Gran2018,Gran2018b,Mauri2018, Romero-Bermudez2018}. However, these studies have invariably neglected the effects of breaking translational symmetry, which we aim to explore in this paper.

Spatial ordering plays an important role in the response properties of strange metals and interesting ordering phenomena of charge and spin occurs in many cases, see \cite{Fradkin2015} and references therein.
Effective long-wavelength hydrodynamical descriptions of electronic spatial ordering are usually very useful because they are the most efficient way to systematically capture the low-energy dynamics of Goldstone excitations \cite{Lubensky_book}. For example, recent descriptions of the collective hydrodynamic behaviour in charge density waves show that indeed the Goldstone bosons are the key ingredients of the low energy  description and control the linear response  \cite{Delacretaz2016,Delacretaz2017a}. 
However, in some cases, like in the pseudo-gap phase, it is believed that some features cannot be captured by these effective low-energy descriptions \cite{Keimer2014}. 
Holography provides another approach as an effective description of strongly coupled matter with two types of degrees of freedom or sectors. From the point of view of the longitudinal response we can call these sectors the zero sound sector, and the ``quantum critical" (QC) continuum. The role of this holographic QC continuum on transport properties has been extensively studied in the literature \cite{ZaanenBook,Hartnoll:2016Rev,Zaanen:2018}. The precise way in which it enters  in the density response of a translationally invariant holographic system was studied in detail in \cite{Romero-Bermudez2018}. Here, we extend this study using an effective model in which we can study the role of the pseudo-phonons associated to the pseudo-spontaneous  translational symmetry breaking. This model provides an effective description  for fluctuating weakly-pinned translational order in which the zero-momentum optical conductivity is qualitatively different from the usual Drude response \cite{Amoretti2018}: 
\begin{equation*}
\sigma(\omega)\simeq \sigma_o-\frac{\rho^2}{\chi_{\pi \pi} }{\Omega-i\omega\over \omega^2-\omega_o^2+i\omega\Omega}\,,
\end{equation*}
where $\rho$ is the charge density, $\chi_{\pi\pi}$ the static momentum susceptibility and $\Omega$ is the phase relaxation rate related to the small mass of the pseudo-Goldstone \cite{Delacretaz2016,Delacretaz2017a}.  The main feature of this formula is that the maximum of $\sigma(\omega)$ is shifted to a finite frequency. 
This behaviour of the conductivity, sometimes referred to as anomalous Drude, has been observed in various materials at large temperatures where they become bad metals. For example,  in Lantanum copper oxides, the development of a dip of the optical conductivity at low frequency has been suggested to indicate the absence of a zero-frequency collective mode, characteristic of bad metals and has been attributed to effects of strong interactions \cite{Hussey2004}. Other cuprates, like ${\rm Bi}_2{\rm Sr}_2{\rm Ca}{\rm Cu}_2{\rm O}_{8+x}$ (Bi-2212) have also been shown to develop this  behaviour for a wide range of doping \cite{Lupi2000,Hwang2007}. It is not yet clear whether this anomalous Drude behaviour with a pinned peak in the conductivity is actually observed in the regime of temperature and doping relevant for the observations of plasmons.\footnote{These measurements are currently being revisited and there is evidence suggesting that in the regime where the plasmon is over-damped, the conductivity in this family of materials (Bi-$2212$) is Drude-like (private communication with  Erik van Heumen and Jan Zaanen).} Here, we limit ourselves to study the plasmon seen in the density response in an effective holographic theory which displays anomalous Drude behaviour of the optical conductivity in some range of temperature.

In this paper we first study in Sec. \ref{sec:conductivity} the electric response at zero momentum in phases where translational symmetry is broken spontaneously and pseudo-spontaneously. The most salient feature, already presented in \cite{Amoretti2018}, is the presence of a pinned peak whose position does not depend monotonically on temperature. We also show the range of parameters where DC conductivity is independent of the pseudo-Goldstone mass; this is the regime in which the symmetry can be said to be broken pseudo-spontaneously.  In Sec. \ref{sec:density}, we focus on the main interest of the paper: the finite momentum density response. The main results can be summarized as follows: at fixed temperature and low frequencies, the neutral density response displays an asymmetric peak which is not due to a single isolated sound mode. This is in contrast with translational invariant setups in which the zero-sound mode dominates the response. As momentum increases, the attenuation (broadening) of this peak decreases at first, and then increases following the usual sound attenuation. We refer to this effect as \textit{momentum broadening inversion}. Similarly, for fixed momentum, increasing temperature also makes the non-Lorentzian peak narrower at first, and then the peak widens  as it becomes dominated by the sound mode. We comment on the similarities with an effective hydrodynamical description which includes the pseudo-Goldstone. We finish by dressing the density response with the Coulomb interaction, which corresponds to gauging the boundary $U(1)$ theory.

\section{A  holographic model with translational (pseudo)-spon\-taneous symmetry breaking}\label{sec:model}
In this paper, we focus on the gravitational model given by Eq. \eqref{eq:action}. This  model has been shown to be useful as a holographic description of the phenomenology of strongly coupled systems with translational order \cite{Amoretti2017,Amoretti2018}. As we explain below, for the choice of couplings done here it is closely related to the holographic Q-lattices model introduced some time ago \cite{Donos2014b,Donos2014e}.

\begin{align}
S&=\int \dd^{4} x \sqrt{-g} \left[\frac{R -2\Lambda}{2\kappa^2}- \frac{1}{2} (\pd \phi)^2 - V(\phi) - \frac{Z(\phi)}{4q^2} F_{\mu \nu} F^{\mu \nu} -\half\sum_{I=1}^{2}Y(\phi)(\pd \psi_I)^2\right]\,,\nonumber\\
&\hspace{1cm}\Lambda=-\frac{3}{2L^2}\,,\  Z=e^{\gamma \phi},\ Y=(1-e^{\phi})^2,\ V=-{4m^2\over \delta^2}+{4m^2\over \delta^2}\cosh(\delta\phi/2)\,.\label{eq:action}
\end{align}
We choose $\gamma=-1/\sqrt{3}$, $\delta=2/\sqrt{3}$ and $m^2L^2=-2$, which fixes the  behaviour of the scalar field near $u\to0$: $\phi\sim \phi_{(s)} u + \phi_{(v)} u^2+\dots$.
Without loss of generality we choose the AdS radius $L=1$, the bulk couplings for gravity $2\kappa^2=1$ and for the gauge field $q=1$. We use the metric ansatz
\begin{align}
\dd s^2 &= \frac{1}{u^2}\left( - Q_{tt}(u) f(u) \dd t^2 + Q_{uu}(u) \frac{\dd u^2}{f(u)} + Q_{xx}(u) (\dd x^2 + \dd y^2) \right)\,, \label{equ:metric}\\
\notag
f &= (1-u)\left( 1 + u + u^2 - \bar \mu^2 u^3 /4 \right)\,,
\end{align}
and solve the background using the DeTurck method as a boundary value problem as explained in Appendix \ref{app:EOMs}.

This model can be regarded as a description dual to a charge density  wave or of a Wigner crystal, in which translational symmetry is broken in one dimension for the first, or all spatial dimensions for the latter \cite{Amoretti2017,Amoretti2018}. The reason is better understood by looking instead at a related model involving two complex scalar fields $\Phi_I = \varphi(u) e^{i \psi_I(x)}$:\footnote{More generally, one can take the kinetic term to be $Y_\Phi(|\Phi_I|)\delta^{IJ}\partial\Phi_I\partial\Phi_J^*$ without changing the important features related to translational symmetry breaking. The connection to the model of Eq. \eqref{eq:action} is most simply seen by taking 	$Y_\Phi(|\Phi_I|)=1/2$, but a similar argument applies for general $Y_\Phi(|\Phi_I|)$ \cite{Donos2014e}.}
\begin{equation}\label{eq:action_complex}
S=\int \dd^{4}x\,\sqrt{-g}\left[R-{\delta^{IJ}\over 2}\partial\Phi_I\partial\Phi_J^*-\frac14Z_\Phi(|\Phi_I|)F^2-V_\Phi(|\Phi_I|)\right],
\end{equation}
where $I,J$ run over the dual boundary theory spatial dimensions. This model is dual to a CFT deformed by the complex scalars $\Phi_I$ and, in addition to the Abelian gauge symmetry: $F=\dd A$, $A\to A+\dd \Lambda$, there are two global $U(1)$ associated to the constant phase rotation of the two complex fields $\Phi_I$. In other words, these global symmetries are related to a shift symmetry of $\psi_I\to \psi_I+c_I$.
Moreover, we are interested in a solution with $\psi_I=\alpha \delta_{Ii}x^i$. In this situation, the shift and translational symmetries are broken to the diagonal group and a consistent gravitational background may be found.\footnote{While the field configuration chosen for $\psi_I$ breaks both of these symmetries, a transformation combining the generators of both of these symmetries leaves the space time invariant. For example, under the transformations
	$\psi^\mu \to\psi^\mu - c^\mu$ (field shift), and  $x^\mu\to x^\mu + \xi^\mu$ (translation)
	the combined transformation for the non-trivial fields is $\psi^I = x^I \to x^I -c^I+\xi^I$ and leaves $\psi_I$ and the rest of the fields invariant if $c^I =  \xi^\mu\delta_\mu^I$ .
	This is the diagonal subgroup of the internal (field shift) and translational symmetries.}
Importantly, the breaking of translational symmetry is explicit or spontaneous depending on whether the field that breaks this symmetry $\Phi_I$ (or $\varphi$) is sourced or not: it is spontaneous if $\varphi_{(s)}{=}0$, and explicit if $\varphi_{(s)}=\lambda\neq0$, where $\varphi(u\to0)\sim \varphi_{(s)}u^{3-\Delta}+\varphi_{(v)}u^{\Delta}+\dots$ and $2\Delta=1+\sqrt{1+4m_\varphi^2}$, $m_\varphi\equiv V_{\Phi}''(\varphi=0)$.

The dynamics of the phase of the complex scalar field: $\psi_I=\alpha \delta_{Ii}x^i+\delta \psi_I e^{-i\omega t+ipx}$ is the key ingredient that reflects the translational symmetry breaking pattern  at the level of linear response. In other words, they encapsulate the (pseudo)-Goldstone dynamics.\footnote{The Goldstone modes can be identified by acting on the background solution with the Lie derivative along a spatial dimension. This leaves all fields invariant except $\psi_I$. Therefore, the dynamics of the Goldstones are given by the linear response fluctuations $\delta \psi_I$.} Moreover, these dynamics may equally be studied by using the model in Eq. \eqref{eq:action}, because near the boundary it is possible to rewrite the action of Eq. \eqref{eq:action_complex} as that of Eq. \eqref{eq:action} by setting $\Phi_I=\phi e^{i\psi_I}$ and expanding for small $\phi$ and $\psi_I$ \cite{Donos2014e,Amoretti2017}.  Therefore, upon consistent choice of the couplings and potentials, both models are equivalent asymptotically in the UV. 
This choice is important. For example, the mapping between both actions is only possible if we choose the coupling $Y(\phi)$ in Eq. \eqref{eq:action} such that it is quadratic for small $\phi$: $Y(\phi\sim0)\sim \phi^2+\dots$. In this case, the term $Y(\phi)\sum_I (\partial \psi_I)^2\sim \phi^2 \sum_I (\partial \psi_I)^2$ is mapped to a term in the expansion of the kinetic term $\delta^{IJ}\partial\Phi_I\partial\Phi_J^*$ with $\Phi_I=\phi e^{i\psi_I}$. If we choose  $Y(\phi\sim0)\sim \OO(\phi^0)+\dots$ instead, there would not such  mapping of $Y(\phi)\sum_I (\partial \psi_I)^2\sim \sum_I (\partial \psi_I)^2$ to any term in the expansion of  $\delta^{IJ}\partial\Phi_I\partial\Phi_J^*$. In this case, the model of Eq. \eqref{eq:action} can only describe the effects of explicit breaking of translational symmetry. 

To summarize, with the model and couplings given in of Eq. \eqref{eq:action}  translations are broken spontaneously if we impose $\phi_{(s)}=0$, or explicitly  if $\phi_{(s)}=\lambda\neq0$, where $\phi(u\sim0)\sim \phi_{(s)} u+\phi_{(v)}u^{2}+\dots$. Therefore, in the limit when $\lambda$ is the smallest scale, translations can be considered to be broken  pseudo-spontaneously.
We note however that, despite the freedom to choose the pattern of translational symmetry breaking, this model does not allow to study the linear response across the transition between a state in which translational symmetry is broken spontaneously or explicitly. The reason is that in these two situations the boundary conditions of the perturbation fields are different and cannot be deformed into each other, see Appendix \ref{app:EOMs}. This suggests that the linear response observables are discontinuous across the transition between a phase with explicit breaking of translations and a phase with spontaneously broken translations. Recently, this observation was also made in related models where the transition may be studied \cite{Donos2019}.

\section{Electric Response at zero momentum}\label{sec:conductivity}
In this section we summarize known results regarding the electric conductivity in phases with spontaneously and explicitly broken translational symmetry \cite{Amoretti2018}. We first study in Sec. \ref{sec:SSB} the spontaneously broken symmetry phase, where a massless Goldstone excitation is present. Then in Sec. \ref{sec:AC_pseudo} we give a small mass to this Goldstone and study the dependence of conductivity on this mass. 
\subsection{Electric response with spontaneously broken translations }\label{sec:SSB}
\begin{figure}[t]
	\centering
	\includegraphics[scale=0.75]{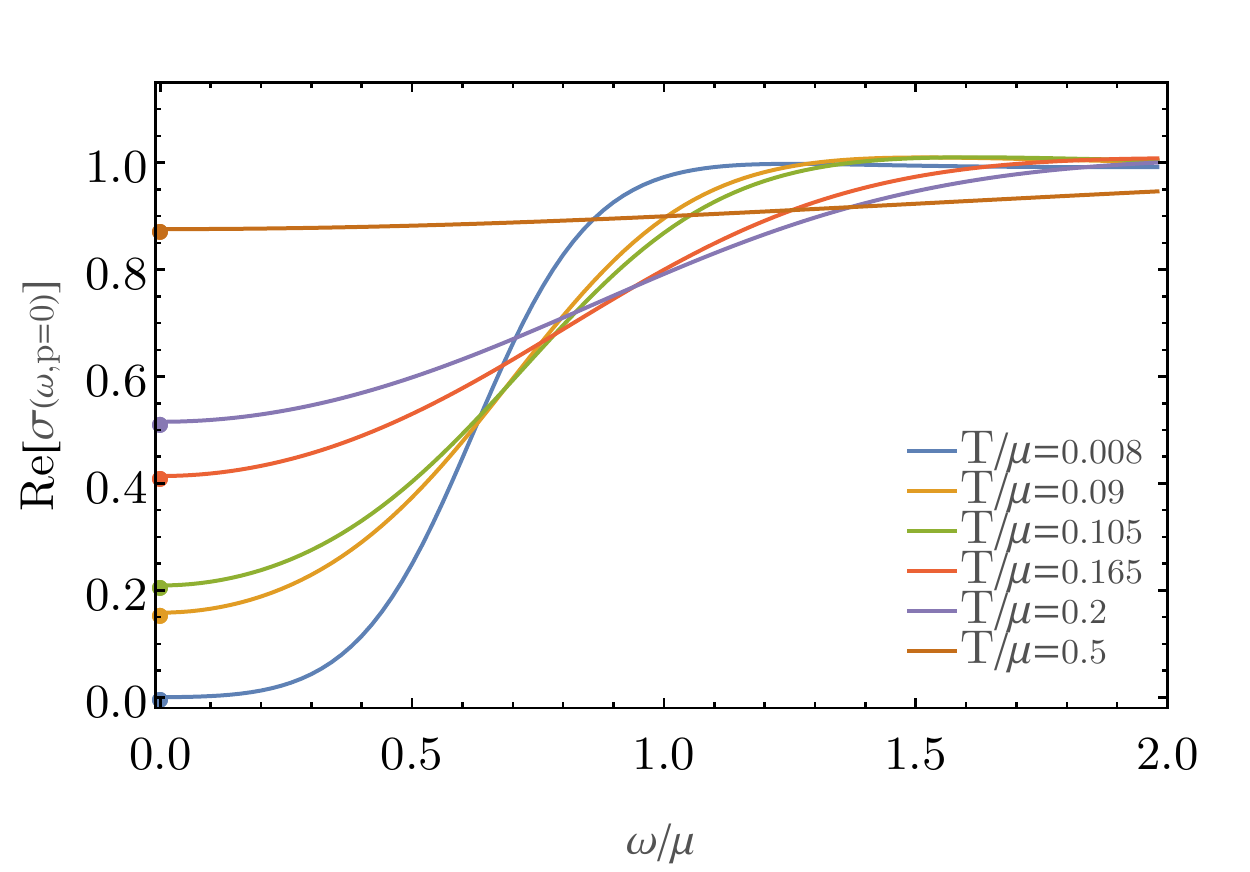}
	\vspace{-2mm}
	\caption{\textbf{Zero-momentum optical conductivity for spontaneously broken translational symmetry}. The dots correspond to the incoherent conductivity computed from Eq. \eqref{eq:DC_SSB}. In addition to this regular part there is an infinite component $K\delta(\omega)$ where the Drude weight is $K=\chi_{JP}^2/\chi_{PP}$. The background is computed with $\lambda/\mu=0$ and $\alpha=10^{-2}$.}\label{fig:AC_Spontaneous}
\end{figure}

When the massive scalar field $\phi$ in the model of Eq. \eqref{eq:action} is not sourced but has a non-trivial VEV, translational symmetry is broken spontaneously giving rise to a gapless fluctuation parametrized by  the massless scalar fields. This gapless (Goldstone) mode plays the role of the phonon. The role of this mode on transport has recently been studied in various holographic constructions \cite{Baggioli2014,Andrade2017,Amoretti2017,Alberte2018,Alberte2018a,Amoretti2018,Donos2019}.
In this phase, the electric DC conductivity has an infinite component parametrized by a Drude weight. The Drude weight  is given by the ratio of susceptibilities $\chi_{JP}^2/\chi_{PP}$ \cite{Amoretti2018}, which saturates the Mazur-Suzuki bound of the Drude weight \cite{Garcia-Garcia2016a}. Moreover, the regular part of the DC electric conductivity at zero momentum is given by \cite{Amoretti2018}
\begin{equation}\label{eq:DC_SSB}
\sigma_{DC}={1\over \chi_{\pi\pi}^2}
\left[Z\big(\phi(u)\big)
\left(sT-\alpha^2\int\limits_0^{u_h} \dd u Y\big(\phi(u)\big) {\sqrt{Q_{tt}Q_{uu}}\over u^2}\right)^2 +{\rho^2\alpha^2\int\limits_0^{u_h} \dd u Y(\phi(u)) {\sqrt{Q_{tt}Q_{uu}}\over u^2} \over Y(\phi(u)) s/4\pi } \right]\,,
\end{equation}
where $u_h=1$, $\rho$ is the charge density and a non-zero value of the parameter $\alpha$ breaks translations $\psi_I=\alpha \delta_{Ii} x^i$.\footnote{Contrary to the incoherent  contribution to the conductivity, which \textit{is} a horizon property, the DC-conductivity Eq. \eqref{eq:DC_SSB} also depends on thermodynamic quantities which in general depend on the bulk details  \cite{Donos2018}. In this case the bulk integral in Eq. \eqref{eq:DC_SSB} is related to the shear modulus as we show below.} Moreover, the entropy density, $s$, scales linearly with temperature at low temperature.  In Fig. \ref{fig:AC_Spontaneous}, we show in dots the DC-conductivity computed from Eq. \eqref{eq:DC_SSB} on top of the numerical  optical conductivity, which increases monotonically with frequency. At low momentum, the longitudinal channel\footnote{To study the modes, we consider a finite-momentum perturbation along the x-direction. The coupled longitudinal modes are specified in Appendix \ref{app:perturbations}.} is dominated by two diffusive poles plus a propagating mode with linear dispersion and an attenuation that increases quadratically with momentum. This hydrodynamic-like attenuation is a consequence of interactions and temperature. In a zero-temperature effective description of a non-Lorentz invariant Goldstone boson, the leading interaction is $(\pd \phi)^3$. Therefore,  Goldstone can decay into two other ones with a decay rate that goes like	  $p^{d+2}$, where $p$ is momentum and $d$ space dimensions. However, in finite-temeprature holographic descriptions, the Goldstone boson couples to other ``thermal degrees of freedom'' resulting in a hydrodynamical decay $p^2$. It would be interesting to explore the Goldstone decay rate in holographic setups with spontaneous breaking of translations at strictly zero temperature. In addition to these three modes, there is also gapped purely imaginary pole, which will eventually collide with one of the diffusive modes for larger momentum.

Moreover, from the holographic renormalization and Ward identities it follows that the shear modulus is 
\begin{equation}\label{eq:shear}
G=\alpha^2\int\limits_0^{u_h} \dd u\  Y(\phi(u))  {\sqrt{Q_{tt}Q_{uu}}\over u^2}+\OO(\alpha^2)\,,
\end{equation} 
which enters, at lowest order, in Eq. \eqref{eq:DC_SSB}, and in the momentum susceptibility: $ \chi_{PP}= sT +\mu\rho-G$. It is also of interest for us the following transport coefficient 
\begin{equation}\label{eq:xi}
\xi = {1\over Y(\phi(u_h))s/4\pi}\int\limits_0^{u_h} \dd u\ Y(\phi(u))  {\sqrt{Q_{tt}Q_{uu}}\over u^2}+\OO(\alpha^2)\,,
\end{equation}
which enters in the diffusive channel of the phonon response at zero momentum $G^R_{\delta\psi_x\delta\psi_x}(\omega) {=} {1\over\chi_{PP}\omega^2}{-}i{\xi\over G\omega}$ \cite{Amoretti2018}. This transport coefficient will become important in the next section when we consider pseudo-spontaneous translational symmetry breaking.

\subsection{Electric response in a holographic Wigner crystal with a light Goldstone}\label{sec:AC_pseudo}

\begin{figure}[t]
 	 		\hspace{-6mm}
 	\begin{subfigure}{0.5\textwidth}
 		\centering{\includegraphics[scale=0.65]{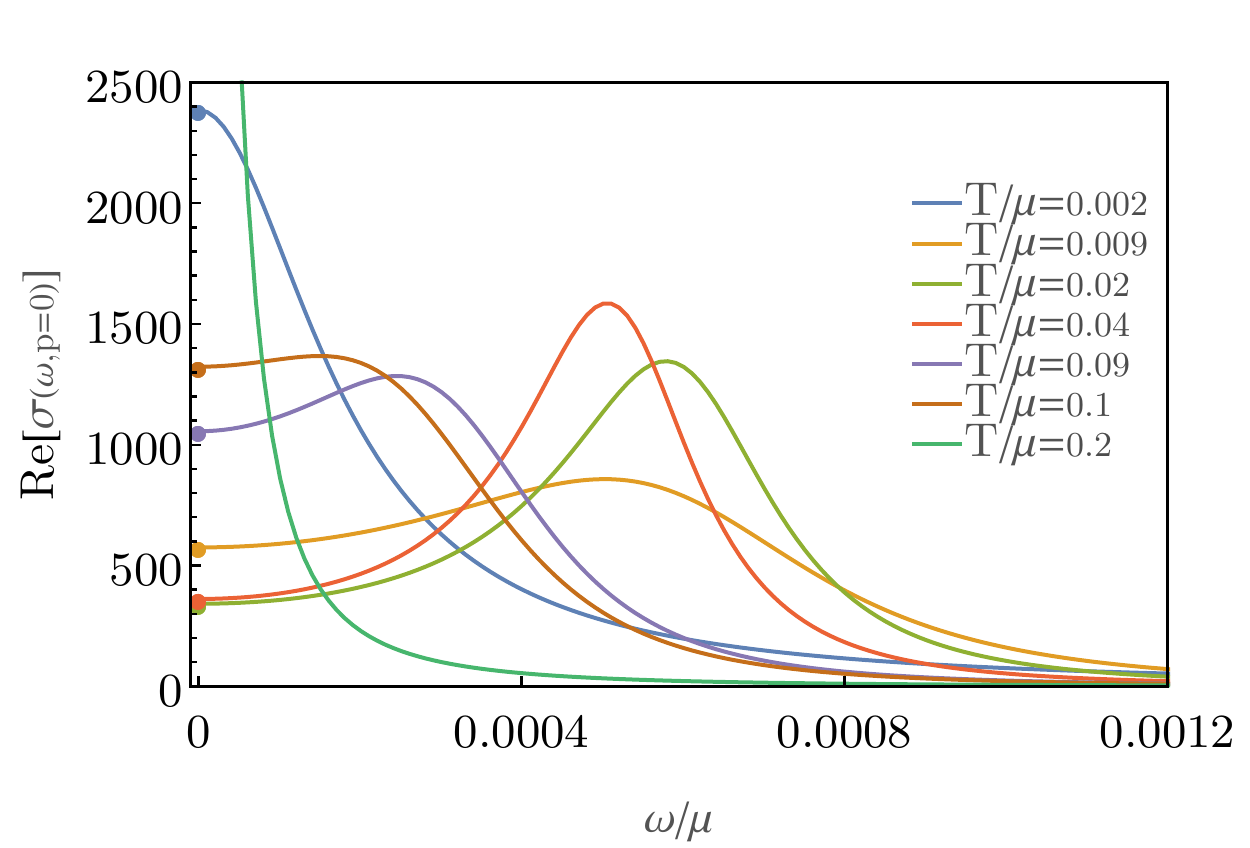}}
 		\caption*{\hspace{1cm}(a)}
 	\end{subfigure}
 	~  \begin{subfigure}{0.5\textwidth}
 		\centering{\includegraphics[scale=0.65]{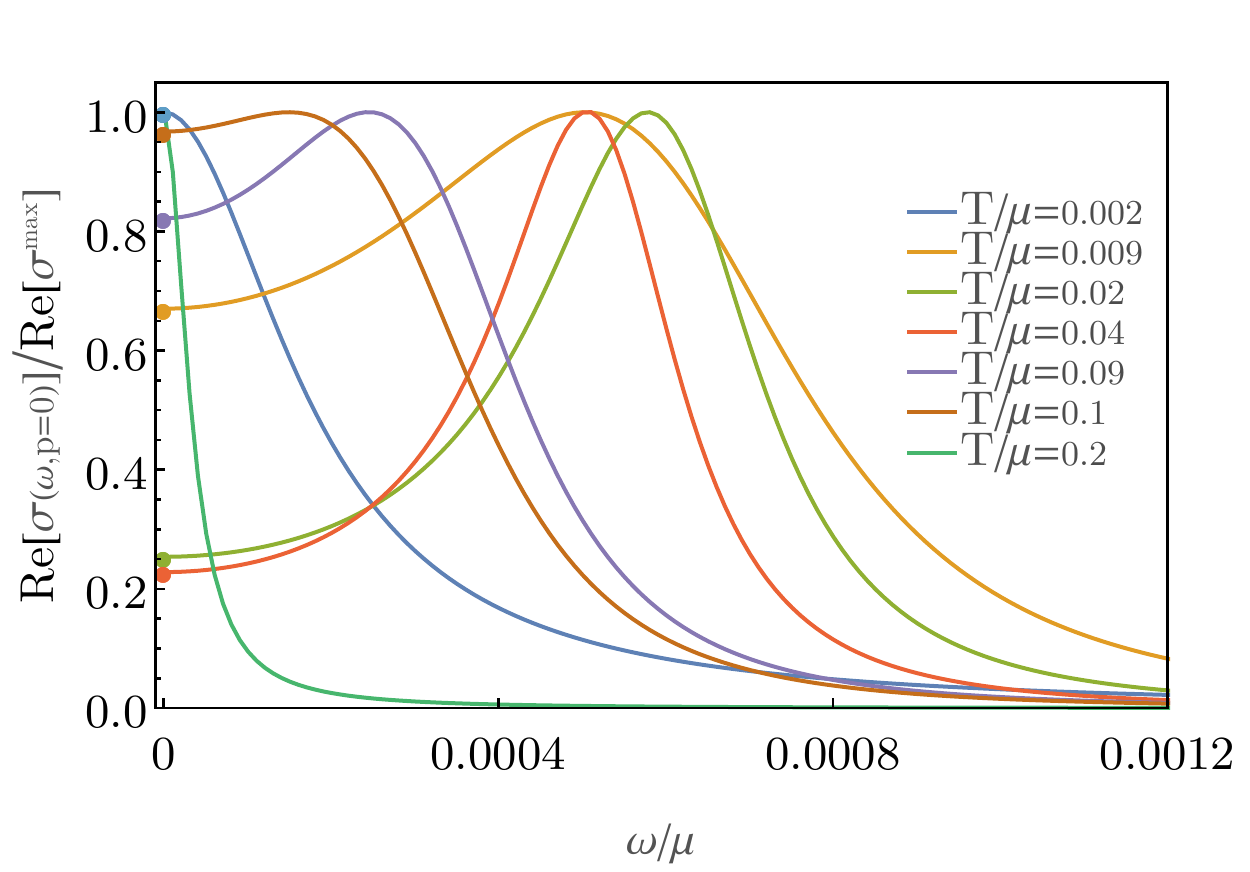}}
 		\caption*{\hspace{1cm}(b)}
 	\end{subfigure}
 	\caption{\textbf{Zero-momentum optical conductivity for pseudo-spontaneously broken translational symmetry}. The dots correspond to the finite DC-conductivity computed from Eq. \eqref{eq:DC_ESB}. We show both the conductivity (left) and the conductivity normalized by the maximum (right). Once the Goldstone associated to spontaneous breaking of translations acquires a light mass, there exists  spectral weight shift from the Drude weight $K\delta(\omega)$ to finite frequencies.  The background is computed with $\lambda/\mu=-10^{-4}$ and $\alpha/\mu=10^{-2}$. }\label{fig:AC_PseudoSpontaneous}
 \end{figure}

As explained in Sec. \ref{sec:model}, when the massive scalar field is sourced with $\lambda\neq0$:
\begin{equation}
\phi(u\to0)\sim \lambda u+\phi_{(v)} u^2+\dots\,,
\end{equation}
 transational symmetry is formally broken explicitly. However, by taking this scale of explicit symmetry breaking to be the smallest scale in the problem, it is possible to retain some of the physics of the (pseudo-) Goldstone mode which now acquires a small mass \cite{Amoretti2018}.
 The gapping of the Goldstone associated to breaking of translations leaves a clear imprint on the electric response, namely the shift of spectral weight from the Drude weight at zero frequency to finite frequencies as shown in Fig. \ref{fig:AC_PseudoSpontaneous} \cite{Delacretaz2016,Delacretaz2017a,Andrade2017,Amoretti2018}.  This is qualitatively similar to the optical measurement in certain cuprate oxides in the bad metallic phase \cite{Hussey2004,Lupi2000,Hwang2007}. We note however that there is an important distinction. Fig. \ref{fig:AC_PseudoSpontaneous} shows that the position of the peak does not increase monotonically with temperature, but instead shifts back to smaller frequencies until eventually the conductivity becomes Drude-like. We can understand this behaviour by looking at the hydrodynamic modes that dominate the longitudinal channel at zero momentum \cite{Delacretaz2016,Delacretaz2017a}
\begin{equation}\label{eq:WC_modes}
\omega_\pm=-i{\Gamma+\Omega\over 2}\pm\half \sqrt{4\omega_o^2-(\Gamma-\Omega)^2}\,,
\end{equation}
where $\Gamma$ and $\Omega$ are the momentum and phase relaxation rates, respectively. The phase relaxation rate originates from the pseudo-Goldstone term which enters in the modified hydrodynamical equation for non-conservation of momentum \cite{Delacretaz2016,Delacretaz2017a}: 
\begin{equation}\label{eq:hydro_mom}
\dot \pi^i +\pd_j \tau^{ij}=-\Gamma \pi^i-G m_{PG}^2\phi_{PG}\,,
\end{equation}
 where $G$ is the shear modulus, $m_{PG}$ is the mass of the pseudo-Goldstone $\phi_{PG}$.
In the holographic model used here,  the phase relaxation rate is given by \cite{Amoretti2018} 
\begin{equation}
\label{OmegaAnalytical}
\Omega^{-1}=\frac1{4\pi T}\int_0^{r_h} dr\left(\frac{4\pi T C_h Y_h \sqrt{B}}{C Y \sqrt{D}}-\frac1{r_h-r}\right).
\end{equation}

\begin{figure}[t]
	\hspace{-5mm}
	\begin{subfigure}{0.5\textwidth}
		\centering{\includegraphics[scale=0.65]{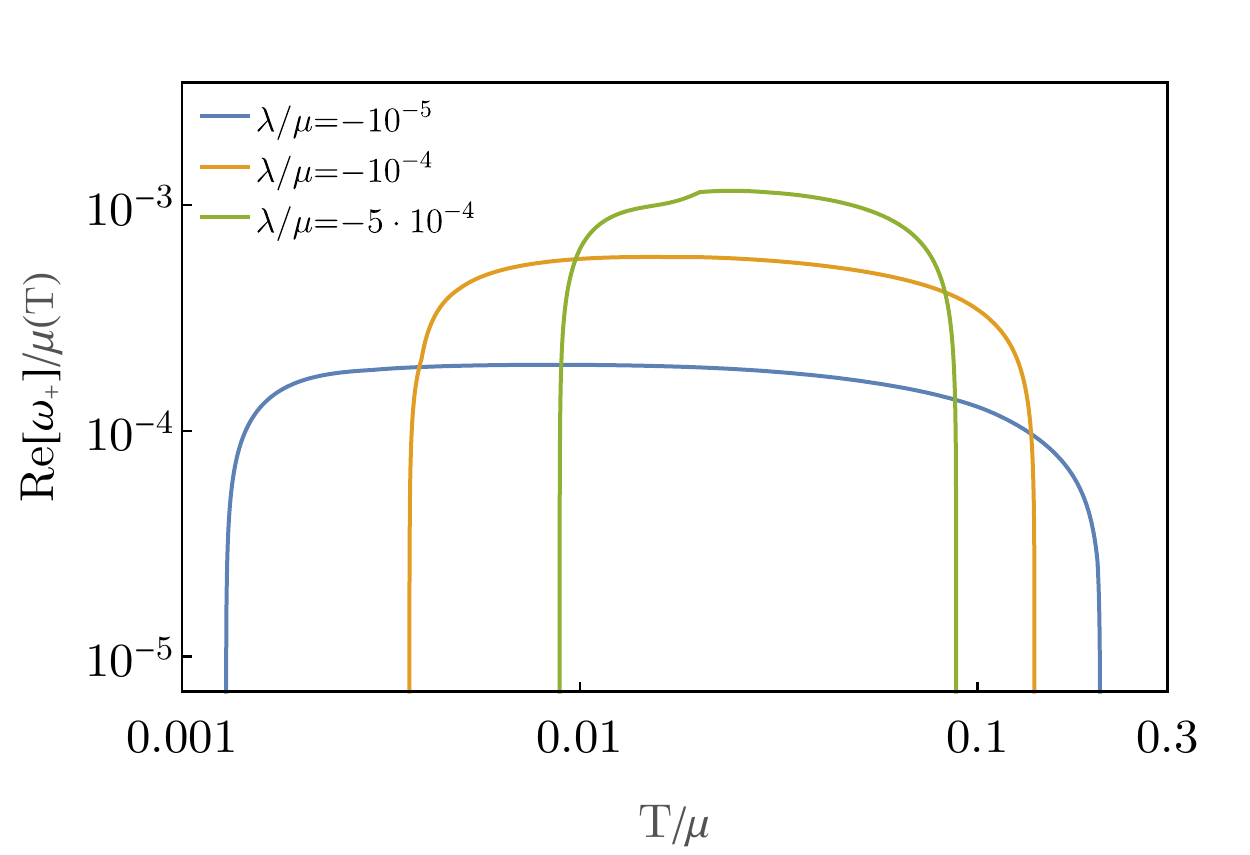}}
 \caption*{\hspace{1cm}(a)}
	\end{subfigure}
	\begin{subfigure}{0.5\textwidth}
		\centering{\includegraphics[scale=0.65]{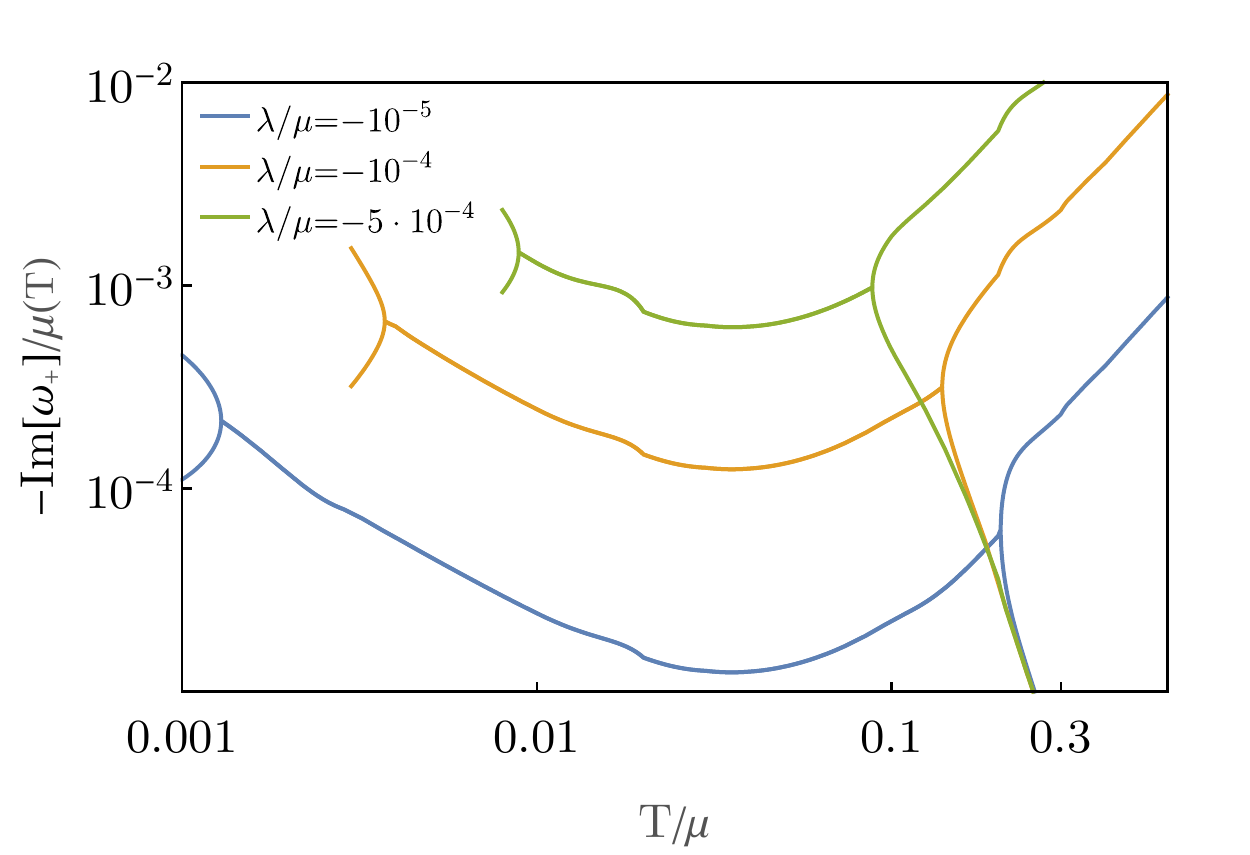}}
 \caption*{\hspace{1cm}(b)}
	\end{subfigure}
	\caption{\textbf{Dominant zero-momentum hydrodynamical modes}, Eq. \eqref{eq:WC_modes}, as a function of temperature for various values of the explicit-symmetry breaking scale $\lambda$ which controls the pseudo-Goldstone mass. The modes have a finite real part only for some range of temperature. This is the range for which the optical conductivity has a pinned peak in Fig. \ref{fig:AC_PseudoSpontaneous}. }\label{fig:qnm_Pseudo_ana}
\end{figure}

Eq. \eqref{eq:WC_modes} shows the hydrodynamic modes have a real part whenever $4\omega_o^2-(\Gamma-\Omega)^2>0$, and are purely imaginary otherwise.  In the first case, the conductivity displays a pinned peak at $\Re(\omega_+)$. On the other hand, when $\omega_{\pm}$ is purely imaginary the optical conductivity will be Drude-like. As shown in Fig. \ref{fig:qnm_Pseudo_ana}, both the temperature and the explicity symmetry breaking scale $\lambda$  control which of these two situations is realized. It is also clear that, for the range of temperatures where $\Re(\omega_\pm)\neq0$, $\Re(\omega_\pm)$ does not depend monotonically on temperature. Moreover, as the scale of explicit symmetry breaking increases the temperature range for which a pinned peak exists shrinks. This is associated to a progressive breakdown of translational symmetry being broken pseudo-spontaneously in favour of an explicit translational symmetry breaking, where the AC-conductivity is Drude-like. 

\begin{figure}[t]
	 \hspace{-5mm}
	\begin{subfigure}{0.5\textwidth}
	   \centering{\includegraphics[scale=0.65]{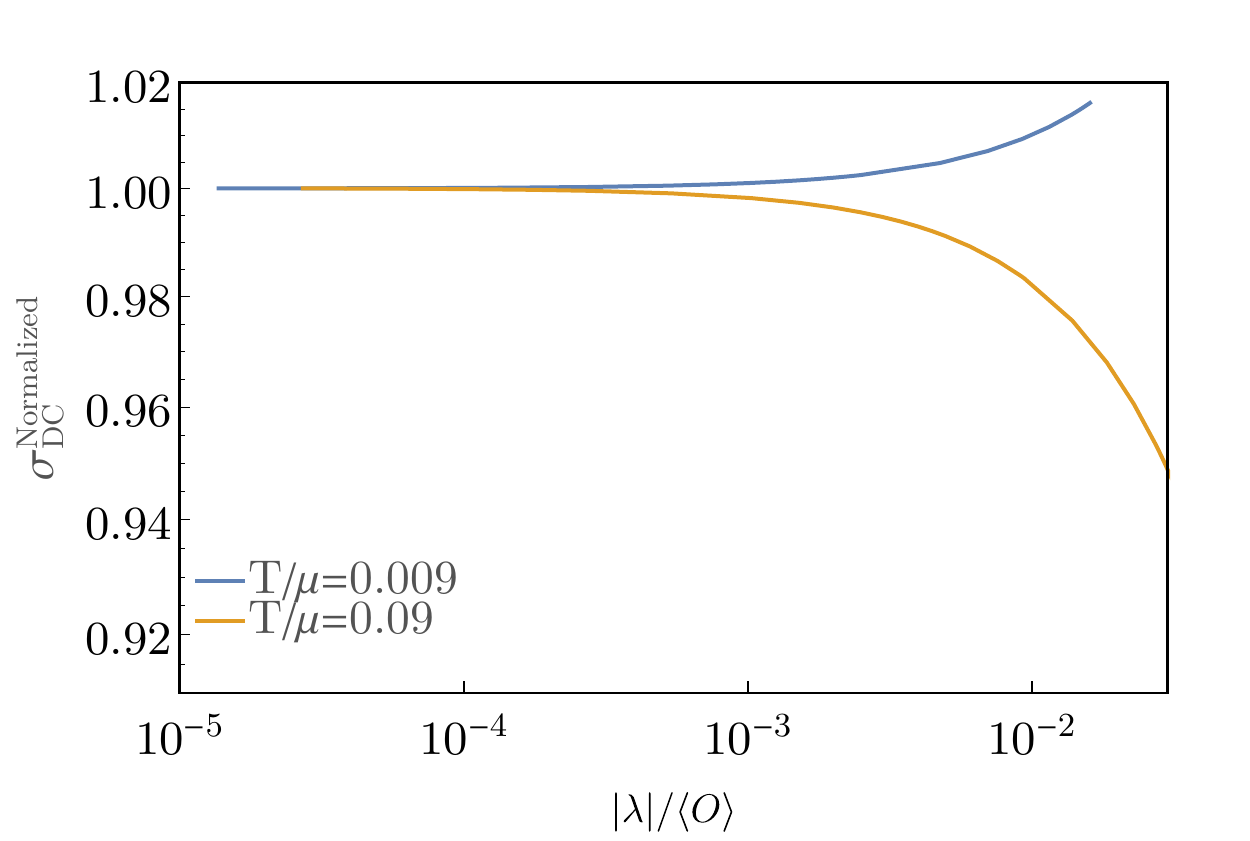}}
		\caption*{\hspace{1cm}(a)}
	\end{subfigure}
	~  \begin{subfigure}{0.5\textwidth}
		\centering{\includegraphics[scale=0.65]{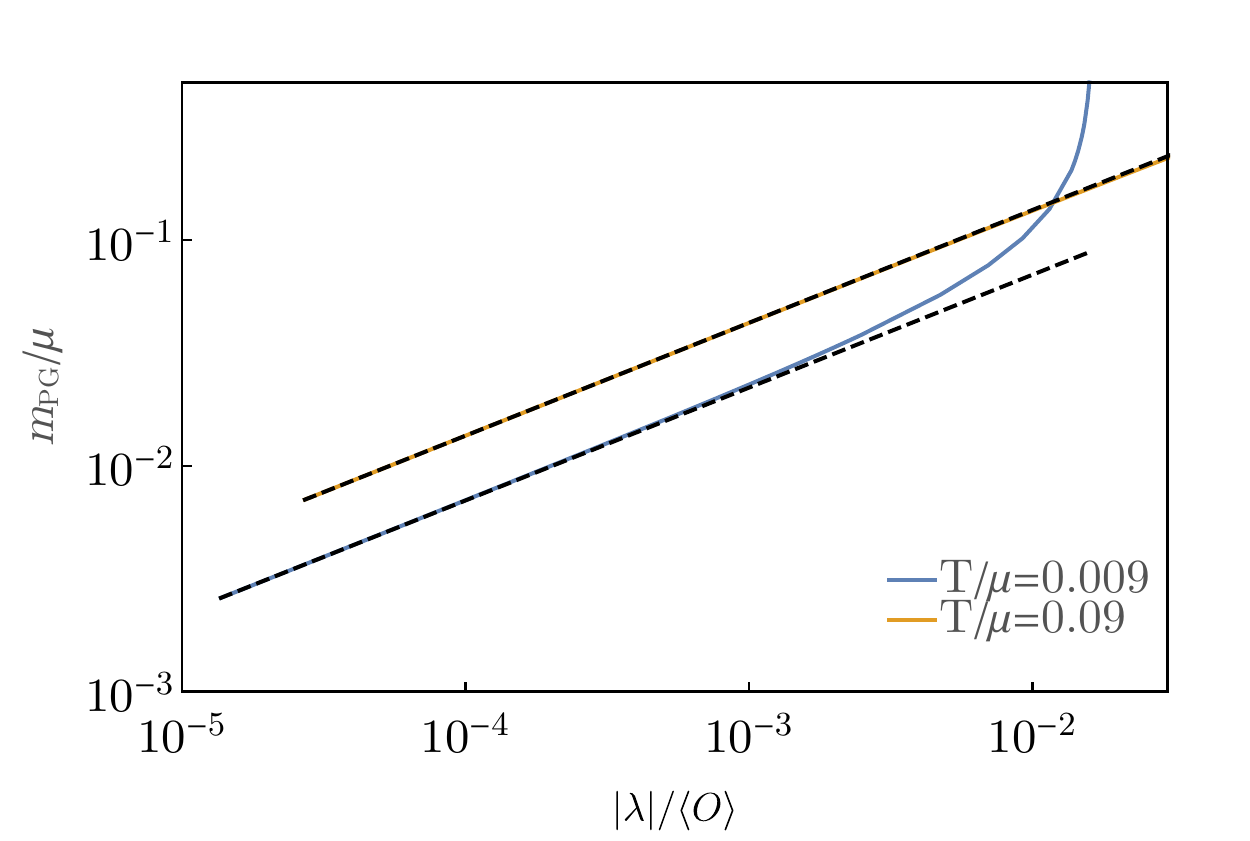}}
		\caption*{\hspace{1cm}(b)}
	\end{subfigure}
\vspace{-3mm}
	\caption{\textbf{Dependence of the DC-conductivity and pseudo-Goldstone mass on the scale of symmetry breaking.} The ratio between the symmetry breaking scale: $|\lambda|$ and the order parameter of $\phi$: $\langle O\rangle \equiv \sqrt{|\phi_{(v)}|}$ controls to what extent translational symmetry is broken explicitly. For small $\lambda/\langle O\rangle $ (and $\lambda/\mu$), the symmetry may be considered to be broken pseudo-spontaneously and the DC-conductivity (left) does not depend on the this scale. In this regime, the pseudo-Goldstone mass (right) increases as $\sqrt{|\lambda|}$ (shown as dashed lines).}\label{fig:DC_Goldstone_mass}
\end{figure}

Contrary to the position of the pinned peak Fig. \ref{fig:qnm_Pseudo_ana}-(a), the DC-conductivity, which is given by 
\begin{equation}\label{eq:DC_ESB}
\sigma_{DC}=
\left.
Z +{\rho^2 \over k^2 Y (\phi{\scriptstyle(u)}) \ s/4\pi }
\right|_{u=u_h}\,,
\end{equation}
is approximately insensitive to the value of $\lambda$, and as a consequence, independent of the pseudo-Goldstone mass, see Fig. \ref{fig:DC_Goldstone_mass}-(a). The region where $\sigma_{DC}$ is approximately constant as a function of $\lambda$ is the region where $\lambda/\langle O\rangle\ll1$  and we can say translational symmetry is broken pseudo-spontaneously. On the other hand, the deviation from this constant value gives an indication when translational symmetry should be considered broken explicitly. 

Moreover, Eq. \eqref{eq:DC_ESB} is valid for any non-zero value of $\lambda$, while Eq. \eqref{eq:DC_SSB} is for strictly spontaneous symmetry breaking $\lambda/\mu=0$. However, in the range of parameters explored $\lambda/\mu\ll1$,  we have observed these two formulas  give drastically different results. The different orders of magnitude of the DC-conductivities in Figs. \ref{fig:AC_Spontaneous} and \ref{fig:AC_PseudoSpontaneous}  suggests that, indeed, there might be a discontinuity in the DC-conductivity at the transition between a phase with spontaneous and pseudo-spontaneous symmetry breaking. This observation has been made in similar models where the transition may be studied explicitly \cite{Andrade2018,Donos2019}. Physically, it is easy to understand that the transfer of spectral weight contained in the Dirac delta at zero frequency in the strict spontaneous symmetry-breaking phase $\lambda=0$ to finite frequencies when  $\lambda\neq0$ will result in a change of the DC-conductivity. However, here we cannot study  this spectral weight shift across the transition between $\lambda=0$ and $\lambda\neq0$.

Finally, we show in Fig. \ref{fig:DC_Goldstone_mass}-(b) the relation between the pseudo-Goldstone mass $m_{PG}$ and the scale of symmetry breaking $\lambda$. In our holographic model, it turns out that the $m_{PG}$  is related to the phase relaxation rate and the diffusive transport coefficient given in Eq. \eqref{eq:xi}  $m_{PG}^2=\Omega/\xi$ \cite{Amoretti2018},
and, for a fixed temperature, it goes as  $|\lambda|^{1/2}$ (dashed lines in Fig. \ref{fig:DC_Goldstone_mass}-(b)). Again, the deviation from the square-root behavior gives a rough estimation of the crossover to a phase where  translational symmetry is broken explicitly.

\section{Response to a density perturbation in a holographic Wigner crystal}\label{sec:density}

In the previous section we studied the longitudinal electric response to a perturbation with zero wave-vector. Here we turn on  a finite wave-vector perturbation and study the density response instead. 
In order to compute the density-density correlation function, we source the density operator $J^0=n(p)$ in the boundary by turning on the boundary value of the temporal component of the $U(1)$ bulk gauge field $\mathcal{A}_0(p)$. The boundary generating functional is then:
\begin{equation}
\label{equ:parti_fun}
\mathcal{Z}[\mathcal{A}] = \int \mathcal{D} \Psi \, \mathrm{exp} \left(-S_\mathrm{boundary}[\Psi] - \int \! dp \, n[\Psi] \mathcal{A}_0 \right).
\end{equation}
As usual in holography the response to this perturbation is then obtained from the decaying modes of the perturbed field:
\begin{align}
\chi^0(\omega,\bp) &\equiv 
\la n(p) n(-p) \ra 
\equiv \left. \frac{\delta^2 \mathcal{Z}[\mathcal{A}]}{\delta \mathcal{A}_0(p) \delta \mathcal{A}_0(-p)} \right|_{\mathcal{A}_0 \rightarrow 0} ={b(p)\over a(p)}\,,\label{eq:neutral_response}
\end{align}
where $A_0(p;u)\Big|_{u\rightarrow0} \sim a(p) \big(1 + \OO(u)\big) + u^{\beta} b(p) \big(1 + O(u)\big) $, for some $\beta>0$. This is the response in a dual theory with a global $U(1)$, i.e., it is the  response in a \textit{neutral} system with conserved density but which does not include the effects of electromagnetism. As explained in detail in \cite{Romero-Bermudez2018}, the charged response is obtained by gauging the boundary $U(1)$. This is achieved by deforming the action with a Coulomb-potential term so that, in the non-relativistic limit, the generating functional is 
\begin{equation}
\label{equ:parti_fun_deformed}
\mathcal{Z}[\mathcal{A}]_{V_{\bp}} = \int \mathcal{D} \Psi \, \mathrm{exp} \left( - S_\mathrm{boundary}[\Psi] - \int \! dp \, n[\Psi] \mathcal{A}_0 +  \int \! dp  \, \frac{1}{2} V_\bp n[\Psi]   n[\Psi] \right)\,,
\end{equation}
where $V_\bp = {e^2\over |\bp|^2}$. In practical terms, the effect of this deformation is to modify the boundary conditions of the time component of the gauge field $A_0(p;u)$. The correct boundary conditions are now  mixed Robin boundary conditions:
\begin{equation}
\label{eq:Robin}
a(p) - e^2 V_\bp b(p) \equiv A_0(p;u)\Big|_{u\rightarrow 0}\hspace{-3mm} - e^2 V_\bp \p_u A_0(p;u)\Big|_{u\rightarrow 0}  \hspace{-3mm}= \mathcal{A}_0(p).
\end{equation}
Physically, this is understood as a redefinition of  the source $\mathcal{A}_0(p)$.  Therefore, the holographic prescription to compute the \textit{dressed}, or charged, response function is modified to \cite{Romero-Bermudez2018}
\begin{align}
\chi(\omega, \bp) &\equiv \la n(p) n(-p) \ra_{V_\bp} = \left. \frac{\delta^2 S_{V_\bp}}{\delta \cA_0 \delta \cA_0} \right|_{\cA_0 \rightarrow 0} = \frac{\chi^0 (\omega,\bp)}{1 -  V_{\bp} \chi^0 (\omega,\bp)},\label{equ:chi_holography}
\end{align}
where $S_{V_{\bp}}$ is the deformed action bulk action: $S_{V_{\bp}} = S_{\rm AdS} {-} \half\int\! \dd^{3}p\, V_\bp b(p) \, b({-}p)$ and $b(p)=\delta \la n(p)\ra$.

\begin{figure}[t]
	\hspace{-0.6cm}
	\begin{subfigure}{0.52\textwidth}
		\centering{\includegraphics[scale=0.6]{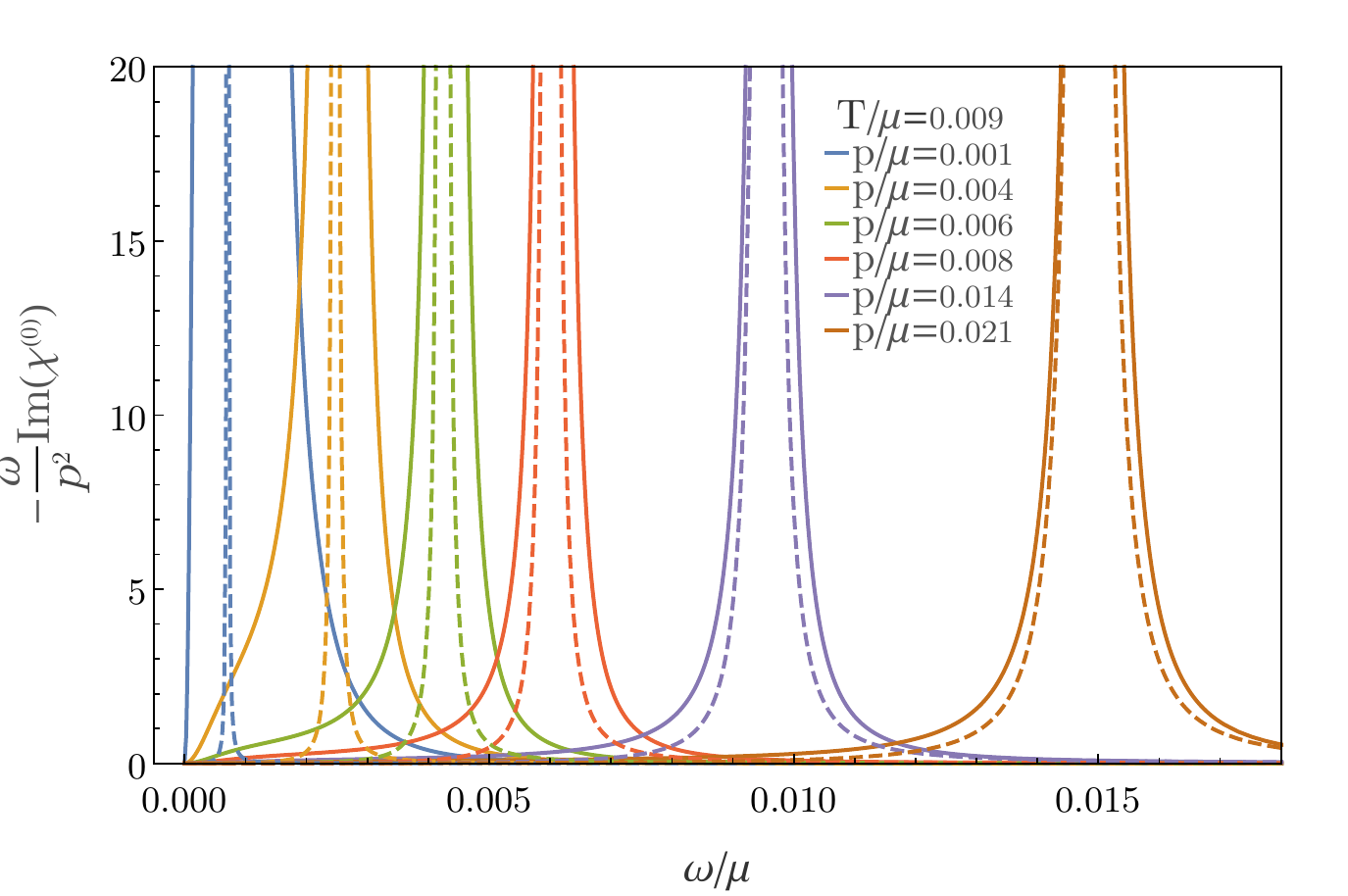}}
		\caption*{\hspace{1cm}(a)}
	\end{subfigure}
	~ \begin{subfigure}{0.52\textwidth}
		\centering{\includegraphics[scale=0.6]{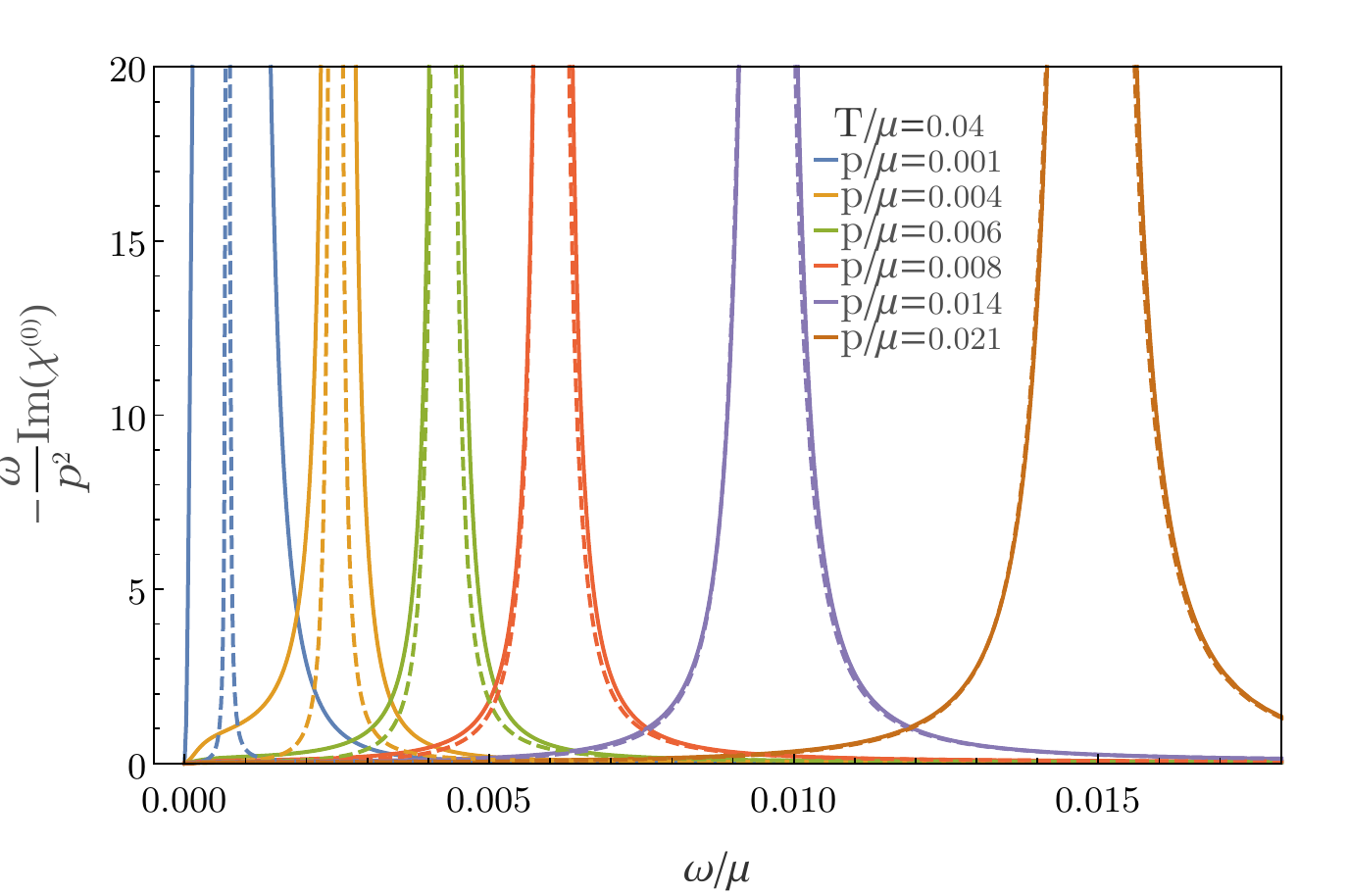}}
		\caption*{\hspace{1cm}(b)}
	\end{subfigure}
	\caption{\textbf{Momentum dependence of the neutral density response}. Continuous  lines: translational symmetry is broken pseudo-spontaneously $\lambda/\mu=-10^{-4}$, $\alpha/\mu=0.1$. Dashed lines: translationally invariant system $\lambda/\mu=\alpha/\mu=0$. When translations are broken, the density response displays a broad asymmetric peak at low momentum. As momentum increases, this peak becomes narrow at first, and then broadens up following the dashed lines. The agreement between dashed and continuous lines indicates the system is insensitive to breaking of translations at large momentum and the attenuation is controlled by the attenuation of the zero-sound. }\label{fig:density_PseudoSpontaneous}
\end{figure}

\subsection{Neutral density response at finite momentum}
We first study the response in a neutral system, Eq. \eqref{eq:neutral_response}, with pseudo-spontaneous translational symmetry breaking. In this section we take the scale of explicit symmetry breaking to be $\lambda/\mu=-10^{-4}$ and $\alpha/\mu=0.1$.  For reference, we compare this density-response with that of a translational invariant system $\alpha=\lambda=0$, analysed in detail in \cite{Romero-Bermudez2018}. The main difference with respect to the latter case is that in the translational invariant system at low temperature the density response is dominated by the sound collective excitation. On the other hand, in the presence of pseudo-spontaneous symmetry breaking at a low, fixed temperature, the density-density response function undergoes a `crossover' from being dominated by a diffusive mode plus a gapped purely imaginary mode to eventually being dominated by the sound mode. The scale of explicit symmetry breaking $\lambda/\mu$  controls the scale where this crossover occurs. The crossover between these two regimes is manifested as a \textit{momentum broadening inversion} shown in Fig. \ref{fig:density_PseudoSpontaneous}. At low momentum, $-\Im\chi^{(0)}$ is dominated by a peak which is clearly not symmetric and this it is not due to a single isolated mode, see continuous $p/\mu=0.001$ lines in Fig. \ref{fig:density_PseudoSpontaneous}. This shape is, in fact, characteristic of two or more nearby purely imaginary poles $\propto -\Im[{1\over (\omega+i\Gamma_1)(\omega+i\Gamma_2)}]$. In contrast, the response in a translationally invariant system  (dashed lines) at the same momentum and temperature is a sharp Lorentzian peak. As momentum increases, the response in the phase with broken translations transitions to a Lorentzian peak and agrees with the data of the translational invariant system, signalling that the density response becomes dominated by the sound mode.

\begin{figure}[t]
	\hspace{-0.5cm}
	\begin{subfigure}{0.52\textwidth}
		\centering{\includegraphics[scale=0.65]{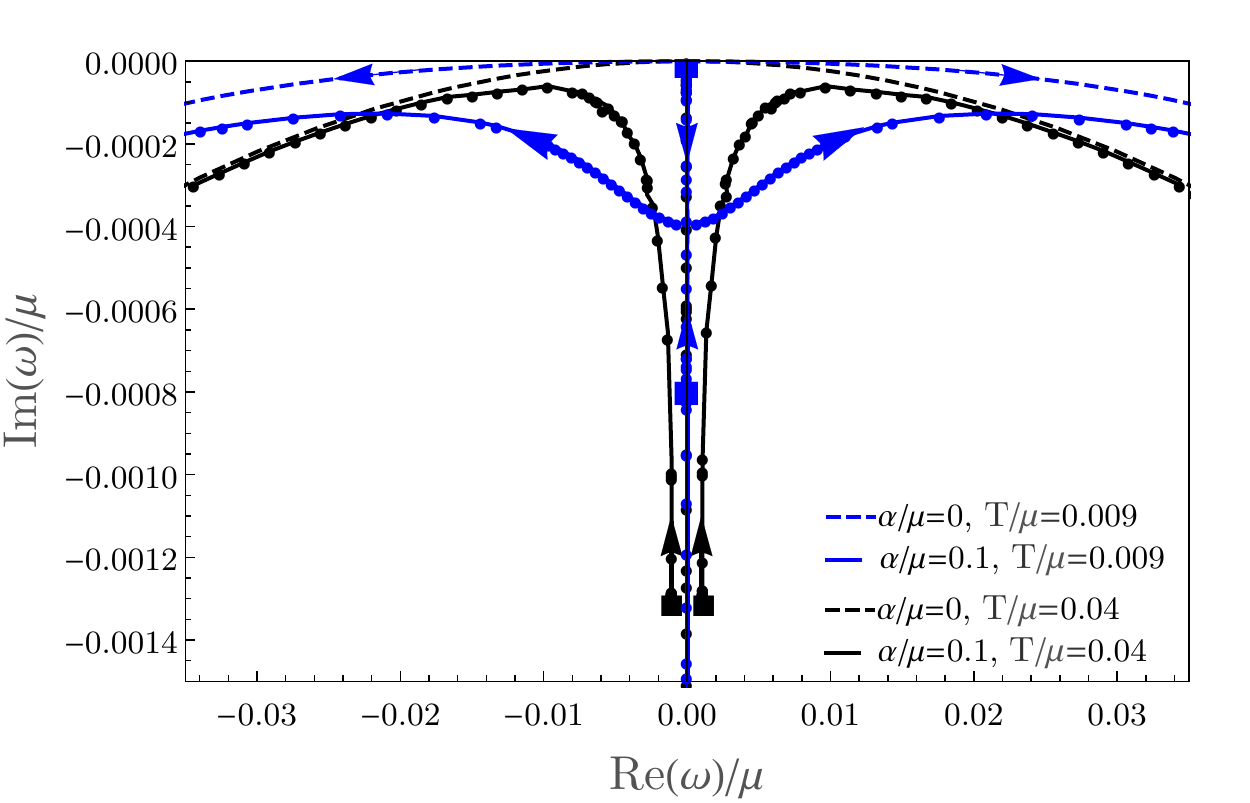}}
		\caption*{\hspace{1cm}(a)}
	\end{subfigure}
	\begin{subfigure}{0.52\textwidth}
		\centering{\includegraphics[scale=0.65]{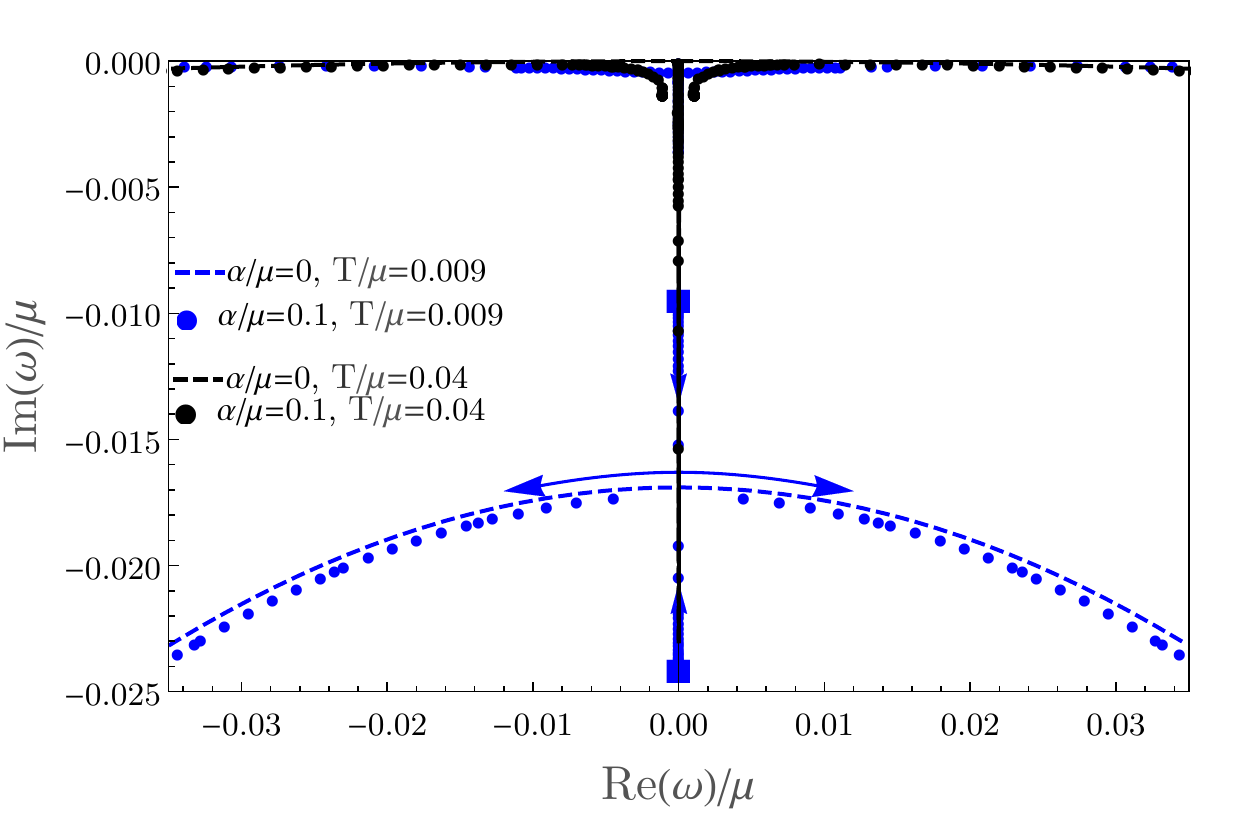}}
		\caption*{\hspace{1cm}(b)}
	\end{subfigure}
	\caption{\textbf{Poles of the density-density correlation function}. Left: Dots and continuous lines:  translational symmetry is broken pseudo-spontaneously with $\lambda/\mu=-10^{-4}$. Dashed lines: zero-sound mode in the presence of translational symmetry.  At low temperature $T/\mu=0.009$, all modes are purely imaginary and two of them collide as momentum increases. For $T/\mu=0.04$, there are two purely imaginary poles (not shown) and the other two poles have a non-zero real part at zero momentum (black squares). In both cases, there is a regime in which the poles move towards the real line. This suggests the density response becomes narrower in this regime. Right plot shows a secondary collision further away from the origin.  }\label{fig:QNMs_PseudoSpontaneous}
\end{figure}

This momentum broadening inversion  is more easily understood from the dynamics of the poles of the  density-density correlator as momentum varies. In Fig. \ref{fig:QNMs_PseudoSpontaneous}, we show the modes that dominate this response. At low momentum, there is not only a single dominant pole near the origin. In fact, the response is dominated by four modes; one ``non-hydrodynamical mode'' that appears at finite momentum plus three hydrodynamical modes which we will analyse in the next section.\footnote{We use the term ``non-hydrodynamical mode'' loosely because it is not included in the three hydrodynamical modes of next section.} At  $T/\mu=0.009$ and $|\bp|/\mu\to0$, these modes are purely imaginary (blue squares in Fig. \ref{fig:QNMs_PseudoSpontaneous}-(a)). 
As momentum increases, two of these poles move closer to each other, collide and approach the real line with an increasing non-zero real part. The location of this collision in the imaginary axis is controlled by the symmetry-breaking scale $\lambda$. Note that this dissipative behaviour is opposite to the usual sound attenuation. Namely,  for some range of momenta, the attenuation of the density response peak decreases as momentum increases. Then, for even larger momentum, these modes eventually become `sound-like' and move down in the complex plane. In this regime, these  poles approach the dashed lines, which correspond the sound mode of the same model but with translational symmetry restored ($\alpha=0$). The behaviour of the imaginary part of these modes is consistent with what we observe in the full density response in Fig. \ref{fig:density_PseudoSpontaneous}-(a): $-\Im\chi^{(0)}$ starts as a broad asymmetric peak at low momentum, and as momentum increases, the peak becomes narrower at first and then broadens up and eventually becomes symmetric at larger momentum.
 
 \begin{figure}[t]
 	\centering{\includegraphics[scale=0.45]{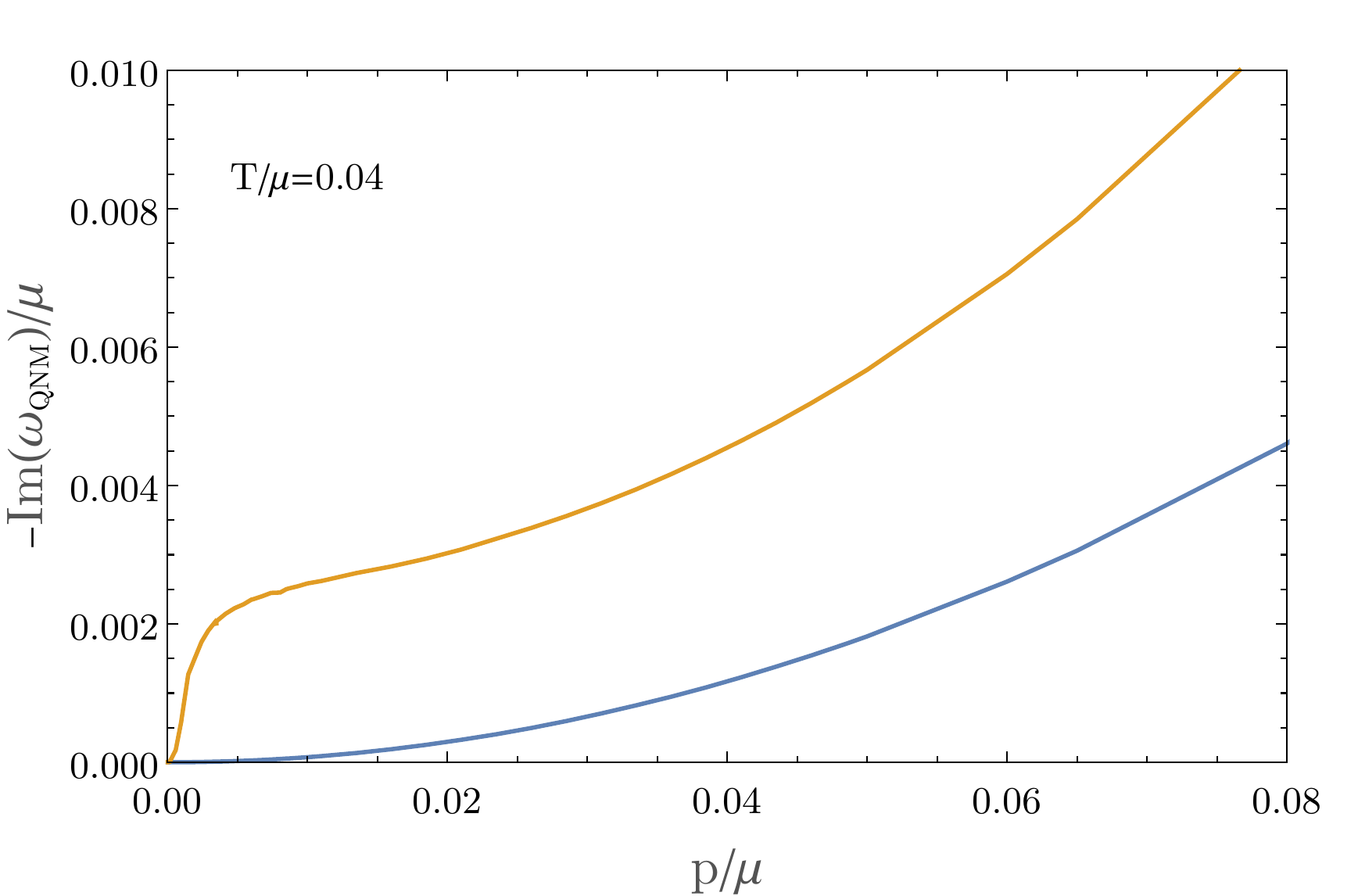}}
 	\vspace{-3mm}
 	\caption{\textbf{Dispersion relation of the two dominant purely imaginary modes}. Blue line corresponds to the diffusion pole. Yellow line shows a non-hydrodynamic mode which disperses down the imaginary axis. At low momentum these two poles dominate the density response shown in Fig. \ref{fig:density_PseudoSpontaneous}-(b).}\label{fig:QNMs_dispersion}
 \end{figure}

At $T/\mu=0.04$ and exactly zero momentum, there are only two purely imaginary modes, which disperse down the imaginary axis as momentum increases. Their dispersion relations are shown in Fig. \ref{fig:QNMs_dispersion}. Contrary to the dynamics at $T/\mu=0.009$,  the other two dominant poles at zero momentum, which are given by Eq. \eqref{eq:WC_modes}, have a small real part, see black squares in Fig. \ref{fig:QNMs_PseudoSpontaneous}-(a). This is because $4\omega_0^2-(\Gamma-\Omega^2)^2>0$ in Eq. \eqref{eq:WC_modes}. As momentum increases, they move up towards the real line and eventually become `sound-like'. However, since their real part is much smaller than their imaginary part at low momentum, these two poles together with the two purely imaginary poles also result into the asymmetric and broad peak in the density response shown in Fig. \ref{fig:density_PseudoSpontaneous}-(b). As before, for larger momentum, the density response peak becomes sharper at first and then broadens up and becomes `sound-dominated' giving rise to a symmetric peak.

Finally, Fig. \ref{fig:QNMs_PseudoSpontaneous}-(b) shows that for $T/\mu=0.009$ (blue), a secondary collision of purely imaginary poles occurs further away from the origin. This collision however, occurs at larger momentum than the collision shown in Fig. \ref{fig:QNMs_PseudoSpontaneous}-(a). In fact, this collision occurs at a scale where the system does not `feel' the effect of translational symmetry breaking, which is reflected by how close the dots (system without translational symmetry) are to the dashed lines (translationally invariant system).

\subsubsection{Hydrodynamical modes with pseudo-broken translations}

Hydrodynamics may be adapted to describe the effects of a light pseudo-Goldstone boson associated to the breaking of translational symmetry. The modification is based on changing the free energy to incorporate the pseudo-Goldstone degrees of freedom. This results on the modification of the momentum ``conservation'' equation with a term that explicitly breaks translations plus a term involving the pseudo-Goldstone mass, Eq. \eqref{eq:hydro_mom} \cite{Delacretaz2017a}.\footnote{An effective field theory based on the coset construction, where the conserved symmetry subgroup is a combination of translations plus internal shift symmetry, has also been presented in \cite{Nicolis2013}.}  Rather than repeating this construction we limit ourselves to study the low-momentum dynamics of these modes. At low momentum the hydrodynamical modes can be obtained as roots of \cite{Delacretaz2017a}:
\begin{equation}\label{eq:hydro_modes}
\omega\big[(\Gamma -i \omega)(\Omega-i\omega)+\omega_0^2 \big] +\omega c^2 \bp^2+i \Omega c_0^2 \bp^2=0\,.
\end{equation}

\begin{figure}[t]
	\begin{subfigure}{\textwidth}
		\hspace{-0.8cm} \includegraphics[scale=1.05]{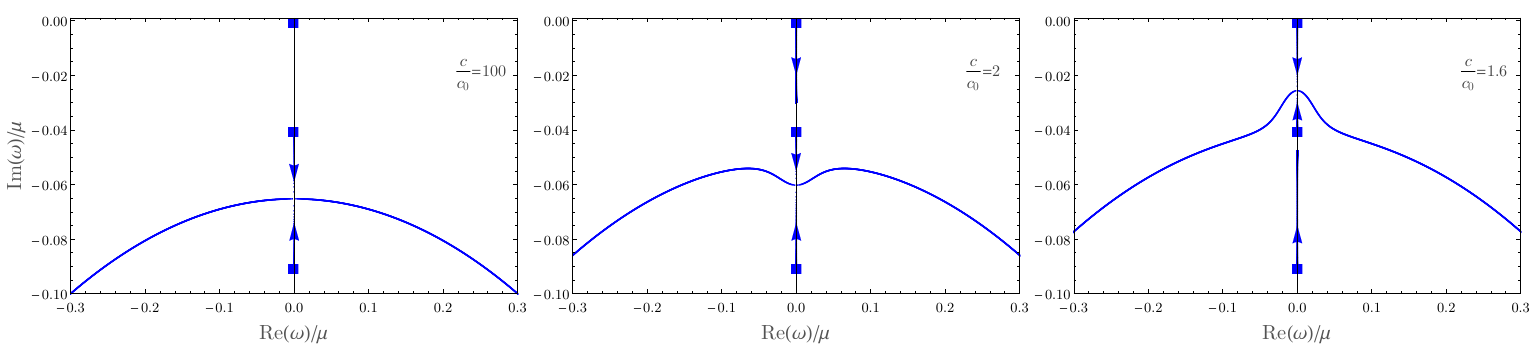}
	\end{subfigure}
	
	\begin{subfigure}{\textwidth}
		\hspace{-0.8cm} \includegraphics[scale=1.05]{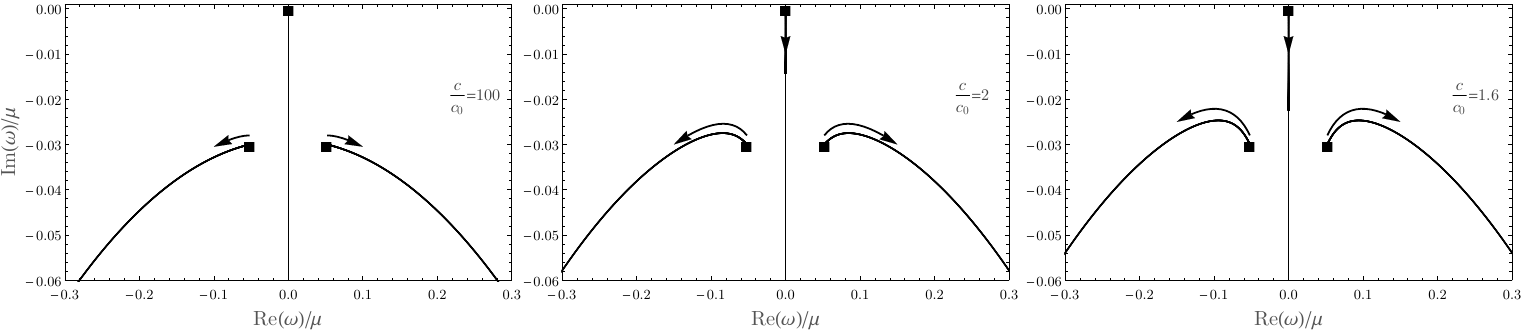}
	\end{subfigure}
	\vspace{-3mm}
	\caption{\textbf{Dispersion of hydrodynamical modes}. Top row is for $\Omega^2-4\omega_0^2>0$ ($\omega_0=0.06$, $\Omega=0.13$) and bottom row for $\Omega^2-4\omega_0^2<0$ ($\omega_0=\Omega=0.06$). When $\Omega^2-4\omega_0^2>0$, all poles are purely imaginary at zero momentum and the dispersion and collision are controlled by the ratio $c/c_0$. For $c/c_0\gtrsim2$ the collision does not involve the diffusive mode while for $c/c_0=1.6$ it does. $c/c_0=2$ is qualitatively similar to that observed in the holographic model for $T/\mu=0.009$ shown in Fig. \ref{fig:QNMs_PseudoSpontaneous}-(a).
		When $\Omega^2-4\omega_0^2<0$, two of the  modes have a non-zero real part at zero momentum and do not collide momentum increases. Their trajectory is again controlled by $c/c_0$. This is the situation observed for $T/\mu=0.04$ in Fig. \ref{fig:QNMs_PseudoSpontaneous}-(a).  The squares correspond to zero momentum: $\omega_{\pm}=\pm\half \sqrt{4\omega_0^2-\Omega^2}-i{\Omega\over 2}$.
		The rest of the parameters are (in arbitrary units) $c=0.25$, $\Gamma=\Gamma_0 p^2$, $\Gamma_0=0.05$. }\label{fig:hydro_QNMs}
\end{figure}

Eq. \eqref{eq:hydro_modes} shows that hydrodynamics predicts three modes.\footnote{In order to make connection to our holographic model we set $\Gamma=\Gamma_0 \bp^2$.}  One of this modes is diffusive and thus is always purely imaginary. Depending on the sign of $\Omega^2-4\omega_0^2$, the other two modes may acquire a real part, even at zero momentum (see Eq. \eqref{eq:WC_modes}): for  $\Omega^2-4\omega_0^2>0$ they are non-degenerate and purely imaginary,  and for 
$\Omega^2-4\omega_0^2<0$ they have non-zero real part. These two scenarios  are indeed seen in the holographic modes shown in Fig. \ref{fig:QNMs_PseudoSpontaneous}, where at $|\bp|=0$ (squares) the modes shown in blue lie on the imaginary axis but the modes shown in black have non-zero real part. For $\Omega^2-4\omega_0^2=0$, the two modes are degenerate and purely imaginary. For non zero momentum, the expressions of the roots of Eq. \eqref{eq:hydro_modes} are too lengthy to analyse explicitly. 
Instead, we show the possible dynamics of the hydrodynamical modes depending on the various parameters in Fig. \ref{fig:hydro_QNMs}.  
Given the parametrization  $\Gamma=\Gamma_0 \bp^2$ it is easy to understand that $\Gamma_0$ controls the slope of the dispersion of the poles. Namely, the trajectories become steeper for larger $\Gamma_0$. The effect of the two speeds of sound is shown in Figs. \ref{fig:hydro_QNMs}, both in the regime where they collide and when they do not. The speed of sound $c_0$ controls the diffusive pole. As expected, in the limit $c/c_0\gg1$ the diffusive pole does not seem to move in the left poles of Fig. \ref{fig:hydro_QNMs}. Moreover, the sound-like poles have a monotonically increasing attenuation with momentum. For smaller values of $c/c_0$ the dynamics of the poles change. When $\Omega^2-4\omega_0^2>0$, there is a collision between two purely imaginary poles. Depending on the precise value of $c/c_0$, this collision may involve the diffusive pole (top right) or not (top centre).\footnote{Although not shown in  Fig. \ref{fig:hydro_QNMs}, when $c/c_0<1$ it is also possible to have a collision involving the diffusive mode, like in top-right of Fig. \ref{fig:hydro_QNMs}, but the trajectory of the resulting dispersive modes is  similar to that of top-centre of Fig. \ref{fig:hydro_QNMs}. } On the other hand, when $\Omega^2-4\omega_0^2<0$, there is no collision and the ratio $c/c_0$ controls the initial dependence of the attenuation on momentum (bottom row).

As explained in the previous section, the dynamics of these quasinormal modes explain the momentum broadening inversion of the peak in the density-density correlator. Indeed, the dynamics seen in the middle plots of each row in Fig. \ref{fig:hydro_QNMs} is qualitatively similar to that seen in Fig. \ref{fig:QNMs_PseudoSpontaneous}-(a). We note however, that in the holographic model, a fourth purely imaginary mode is present, see yellow line in Fig. \ref{fig:QNMs_dispersion}. Therefore, even when $\Omega^2-4\omega_0^2<0$, the density-density correlator at low momentum is given by an asymmetric  peak dominated by two nearby purely imaginary poles.

\subsubsection{Thermal broadening}
\begin{figure}[t]
\hspace{-0.5cm}
  \begin{subfigure}{0.52\textwidth}
    \centering{\includegraphics[scale=0.58]{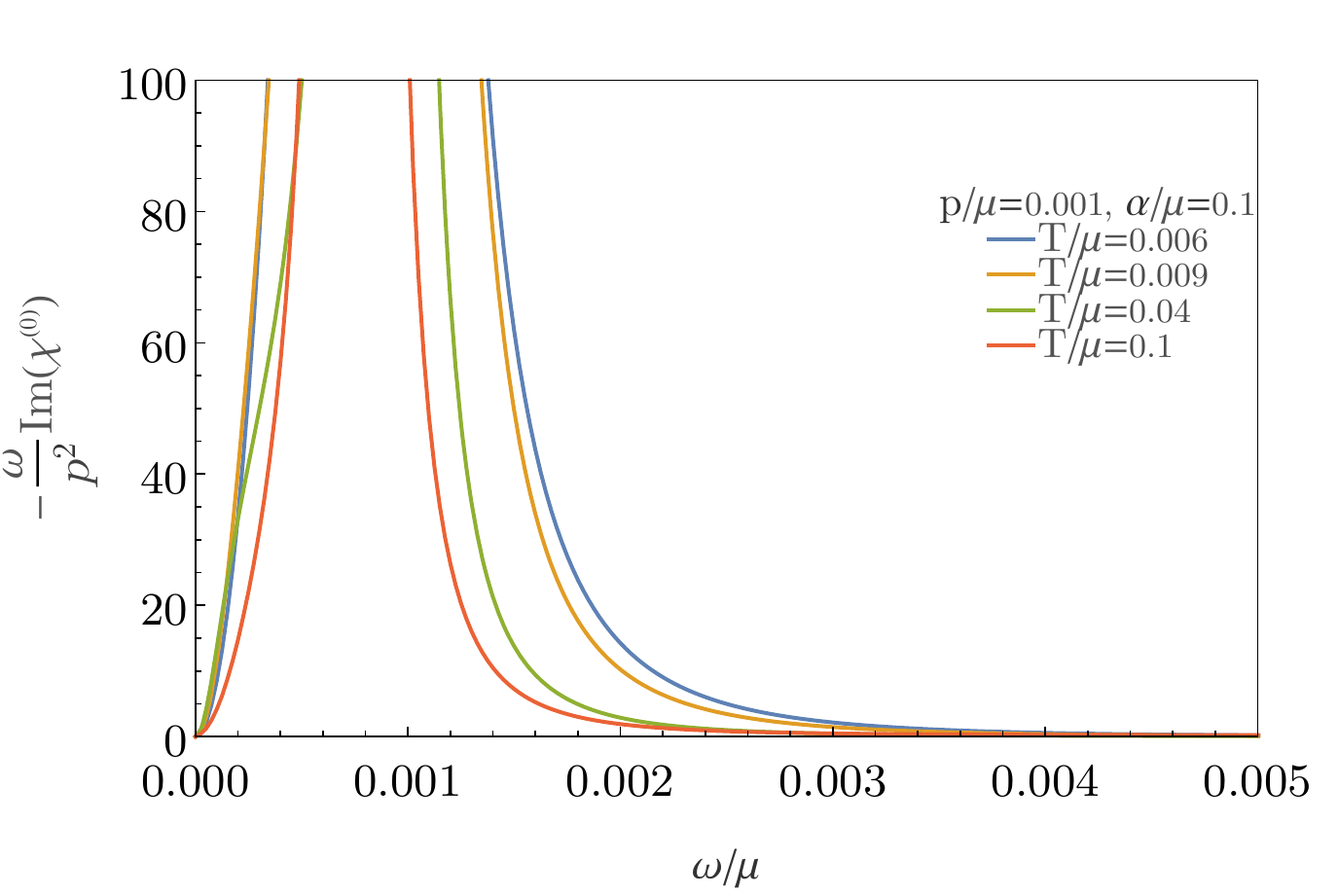}}
     \caption*{\hspace{1cm}(a)}
\end{subfigure}
  \begin{subfigure}{0.52\textwidth}
    \centering{\includegraphics[scale=0.58]{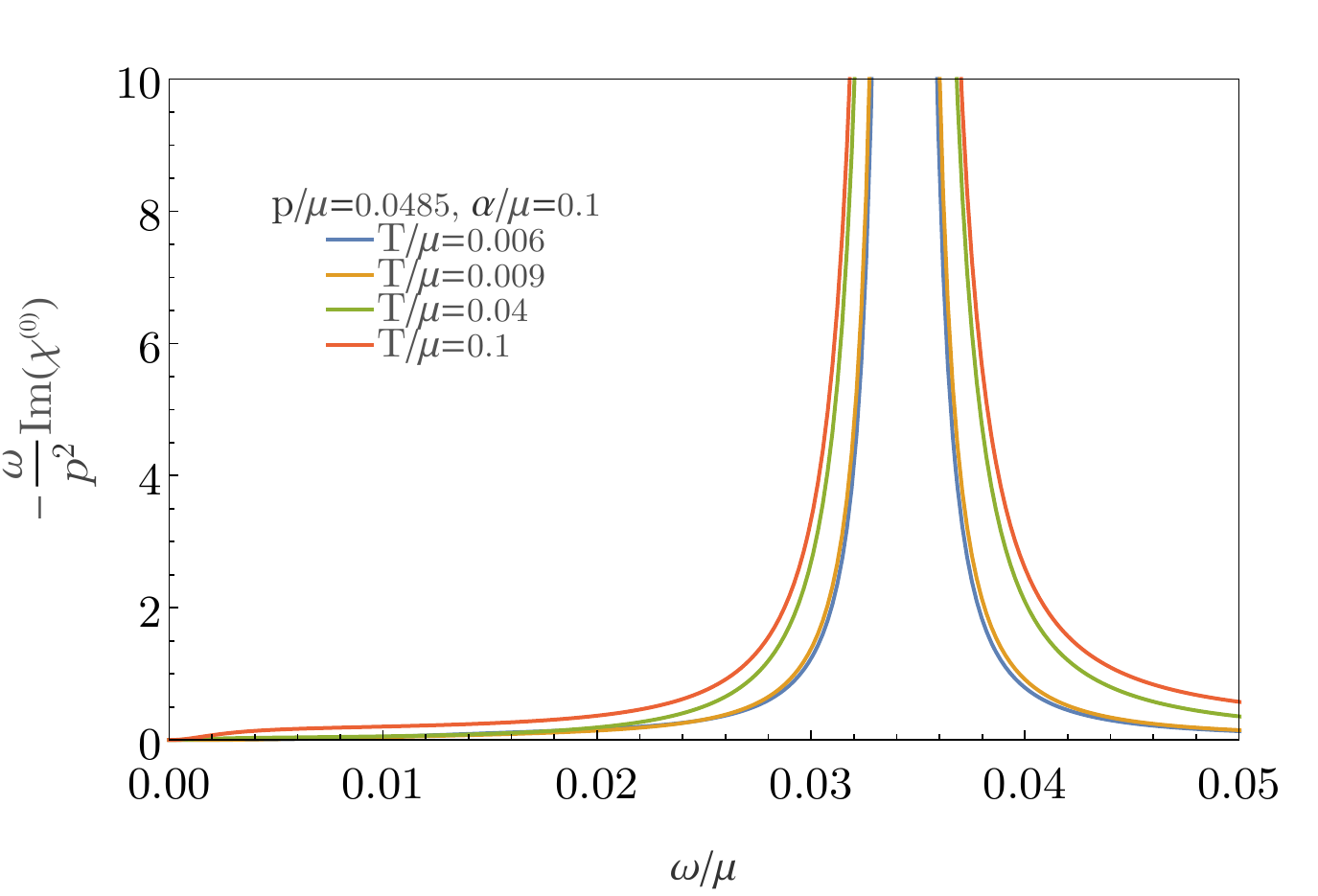}}
    \caption*{\hspace{1cm}(b)}
  \end{subfigure}
\caption{\textbf{Temperature dependence of the neutral density response}. At low momentum the density response becomes longer lived for larger temperature, while at higer temperatures the attenuation increases with temperature. This effect is similar to that observed in Fig. \ref{fig:density_PseudoSpontaneous}, where initially the peak narrows as momentum increases. }\label{fig:neutral_Ts}
\end{figure}

In this section, we show that the effect of temperature on the density response is similar to the effect of momentum shown in the previous section. More specifically, we observe a \textit{thermal broadening inversion}, i.e., instead of the expected increase of the damping rate as temperature increases, we show the existence of a range of temperature in which this damping rate decreases as temperature increases. Beyond this range of temperature, the general expectation of stronger damping for larger temperature is recovered. This effect is nothing but the finite-momentum generalisation of the phenomenon displayed in the zero-momentum optical conductivity of Sec. \ref{sec:AC_pseudo}, where the position of the pinned peak does not depend monotonically on temperature. The thermal broadening inversion is observed at low momentum, Fig. \ref{fig:neutral_Ts}-(a), while for larger momentum, Fig. \ref{fig:neutral_Ts}-(b), the damping rate increases with temperature. This is again due to the interplay between the different scales: the explicit symmetry breaking scale $\lambda$ (that controls the pseudo-Goldstone mass), temperature and momentum. At low temperature and momentum, multiple poles contribute in a similar way to the response but have different dispersion relations. On the other hand, at large momentum or temperature, a sound-like  mode with the usual dispersion dominates the response and the system behaves similarly to the translationally invariant system studied in \cite{Romero-Bermudez2018}. Similar  effects due to temperature have been observed in  the dynamics of the quasinormal modes in related models \cite{Alberte2018,Andrade2018}.

\subsection{Density response dressed with the Coulomb interaction}

\begin{figure}[t]
	\hspace{-0.5cm}
	\begin{subfigure}{0.52\textwidth}
		\centering{\includegraphics[scale=0.65]{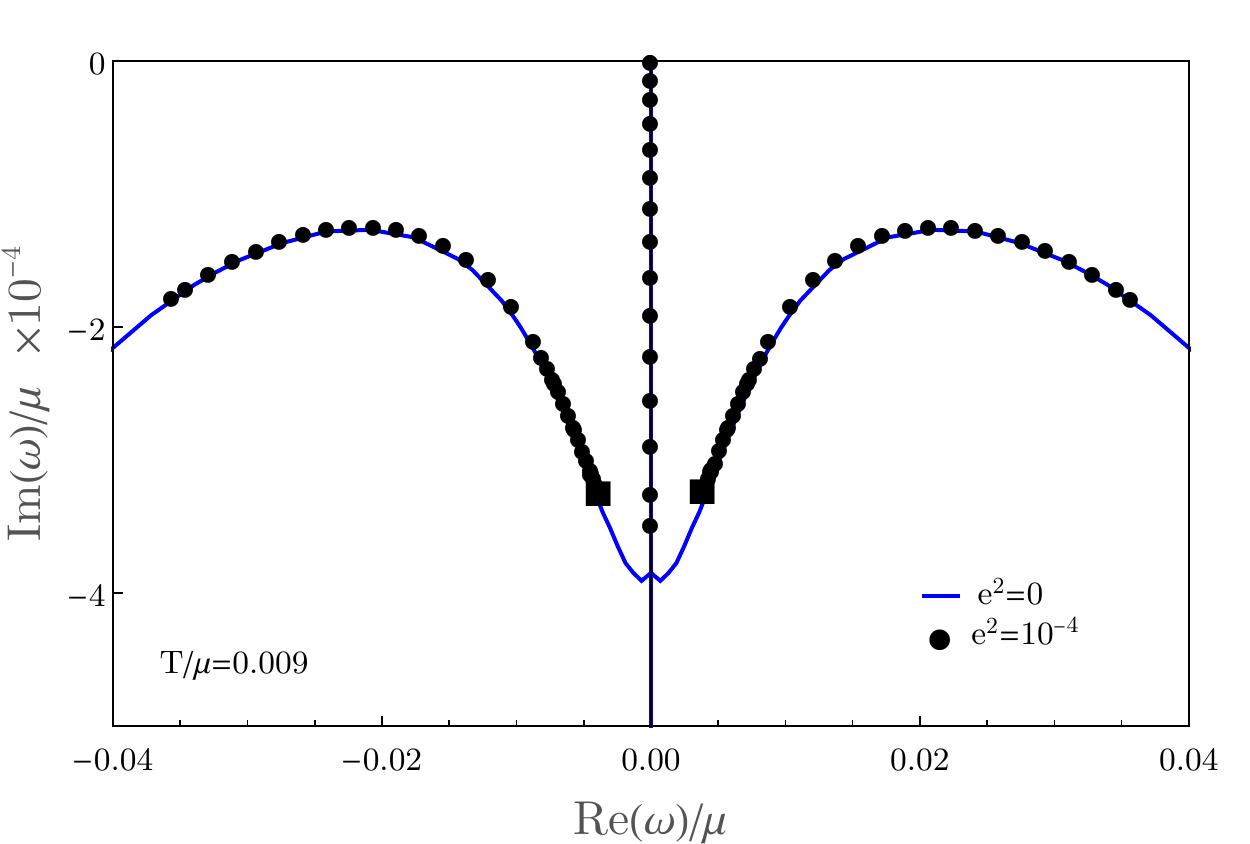}}
		\caption*{\hspace{1cm}(a)}
	\end{subfigure}
	\begin{subfigure}{0.52\textwidth}
		\centering{\includegraphics[scale=0.65]{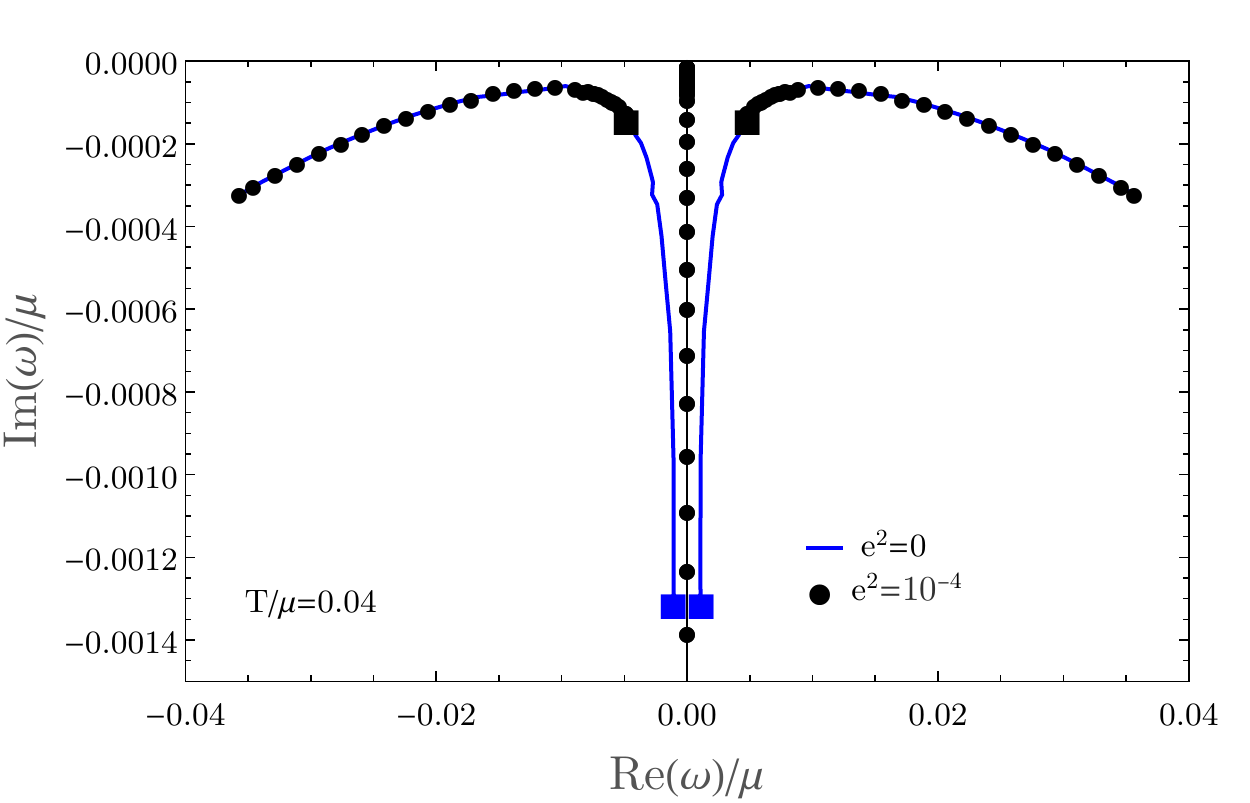}}
		\caption*{\hspace{1cm}(b)}
	\end{subfigure}
	\caption{\textbf{Poles (black) of the dressed density-density correlation function as a function of momentum}. Lines: trajectories of the naked quasinormal modes from Fig. \ref{fig:QNMs_PseudoSpontaneous}. Dots: dressed quasinormal modes computed with mixed boundary conditions Eq. \eqref{eq:Robin}. The effect of turning on the boundary Coulomb interaction is to gap the zero-momentum quasinormal modes, which then follow the same trajectory of the quasinormal modes in the absence of this interaction. }\label{fig:Dressed_QNMs}
\end{figure}
We now study how the density response is modified by gauging the boundary $U(1)$ global symmetry. As explained in the beginning of the section this is achieved by deforming the theory as in Eq. \eqref{equ:parti_fun_deformed} \cite{Romero-Bermudez2018}. Such deformation changes the boundary conditions and the  holographic prescription to compute the two-point function, see Eq. \eqref{equ:chi_holography}. 

As explained in \cite{Romero-Bermudez2018}, for a translationally invariant system, the presence of a sound excitation in the neutral response implies the existence of a gapped mode in the dressed response. This follows immediately from Eq. \eqref{equ:chi_holography}, together with the general parametrization of the neutral response shown in Eq. \eqref{equ:zero_sound}.
\begin{equation}
\label{equ:zero_sound}
{\chi^{(0)}(\omega, \bp)} = \frac{\bp^2 A}{\omega^2 - (v_s \bp)^2 + i \omega \Gamma(\bp) + \Gamma (\bp)^2/4} + \bp^2\Xi(\omega, \bp)\,,
\end{equation}
 where $v_s$ is the speed of sound, $\Gamma$ is the sound attenuation, $A$ is the residue of the sound pole and $\Xi$ encapsulates the quantum critical sector of the theory. Substituting Eq. \eqref{equ:zero_sound} into \eqref{equ:chi_holography} and expanding for low momentum it is easy to see the dressed response is indeed gapped and the attenuation in the $|\bp|\to0$ limit is finite and controlled by the QC continuum sector $\Xi(\omega,\bp)$:
 \begin{align}
\label{eq:chi_dressed}
\hspace{-.3in}
\chi(\omega,\bp)& \simeq \frac{\bp^2 \tilde{A}}{\omega^2-(v_s \bp)^2  - \tilde{\omega}_p^2 + i \omega \tilde{\Gamma}}\\
\tilde{\omega}_p^2 & =  
{A}V_{\bp}\bp^2 +AV_{\bp}^2\bp^4\text{Re}(\Xi)+ \ldots, \nonumber\\
\tilde{\Gamma} &
= \Gamma 
+ AV_{\bp}^2\bp^4 (-\text{Im}(\Xi)/\omega)+\ldots, 
\nonumber \\
\tilde{A} &
 = A + AV_{\bp}\bp^2\Xi +\ldots\ .\nonumber
\end{align}
 However, as shown in Figs. \ref{fig:density_PseudoSpontaneous} and \ref{fig:QNMs_PseudoSpontaneous}-(a), when translational symmetry is broken pseudo-spontaneously, the density response at low temperatures and momentum is not controlled by a single isolated sound pole, but instead by various nearby poles. Therefore, in this case it is not very illuminating to substitute a complicated parametrization for the neutral response with four poles plus $\Xi$ into Eq. \eqref{equ:chi_holography} to obtain   the pole structure of the dressed response $\chi$. Instead, we show directly the effect of including the boundary Coulomb interaction in the two representative cases shown in Fig. \ref{fig:density_PseudoSpontaneous}-(a). 
 Figure  \ref{fig:Dressed_QNMs} shows that effect of turning on the boundary Coulomb interaction, controlled by $e^2$, is to move the two purely imaginary colliding poles at $T/\mu=0.009$  (blue dots in Fig. \ref{fig:QNMs_PseudoSpontaneous}) off the imaginary axis.  These to poles now acquire a finite real part in the $|\bp|\to0$  limit and lie precisely on the same trajectory that the colliding poles of Fig. \ref{fig:QNMs_PseudoSpontaneous} were following; for clarity we show this trajectory as a continuous line also in Fig. \ref{fig:Dressed_QNMs}-(a).  Similarly, the two poles that were further away from the origin at $|\bp|=0$, $T/\mu=0.04$ in the absence of boundary Coulomb interaction (blue squares in Fig. \ref{fig:Dressed_QNMs}-(b)) have now been shifted  closer to the origin (black squares in Fig. \ref{fig:Dressed_QNMs}-(b)) and again follow the same trajectory (blue line) of the poles of the neutral response as momentum increases. Therefore, we see that when the boundary Coulomb interaction is turned on $e^2\neq0$, there are two gapped plasmon modes which follow the expected $\sqrt{\omega_p^2+v^2 \bp^2}$ dispersion. This happens regardless of whether the corresponding naked modes $e^2=0$ are purely imaginary (colliding modes like in the top row Fig. \ref{fig:hydro_QNMs}) or not (bottom row Fig. \ref{fig:hydro_QNMs}).

\begin{figure}[t]
	\hspace{-0.6cm}
	\begin{subfigure}{0.52\textwidth}
		\centering{\includegraphics[scale=0.59]{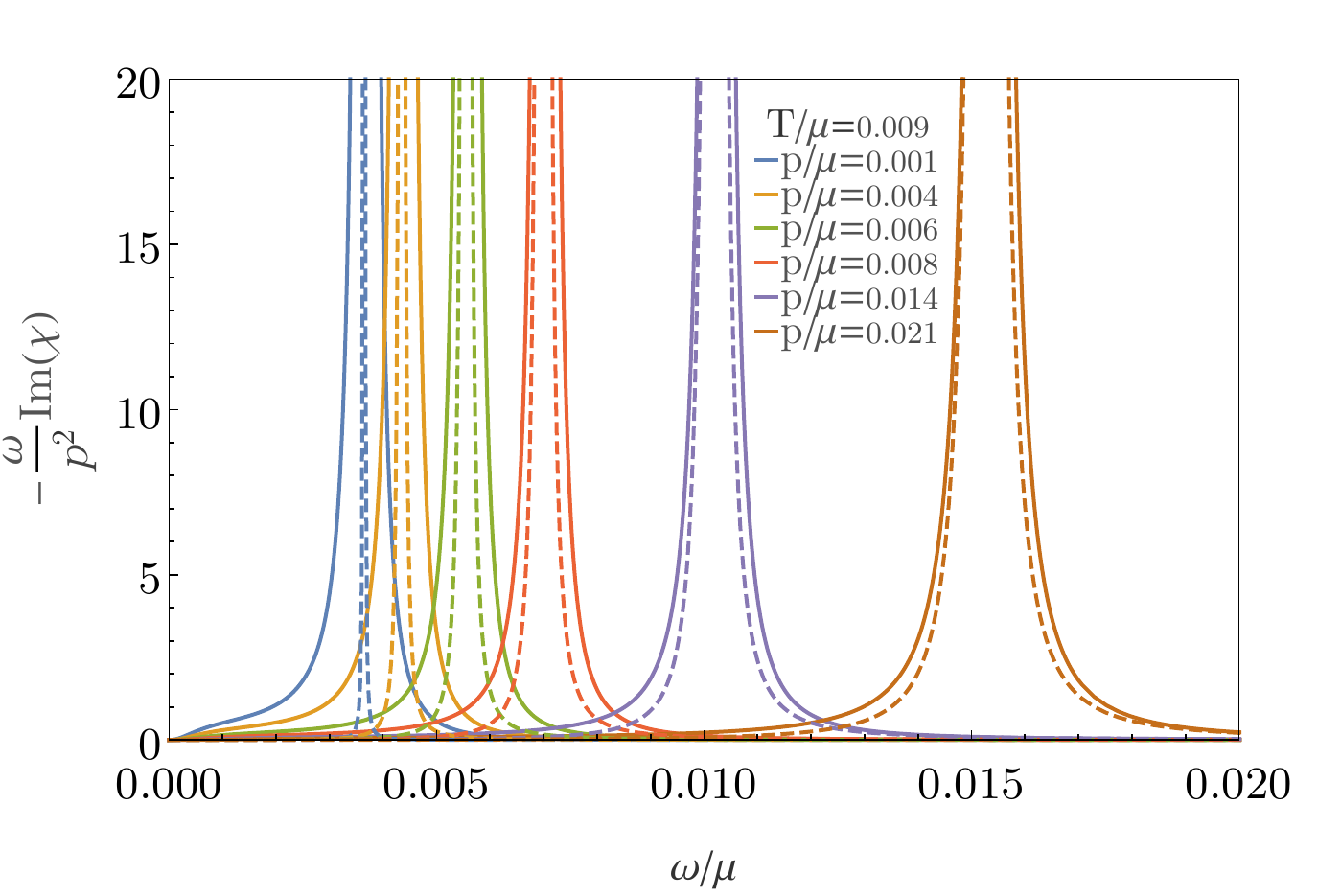}}
	\end{subfigure}
	\begin{subfigure}{0.52\textwidth}
		\centering{\includegraphics[scale=0.59]{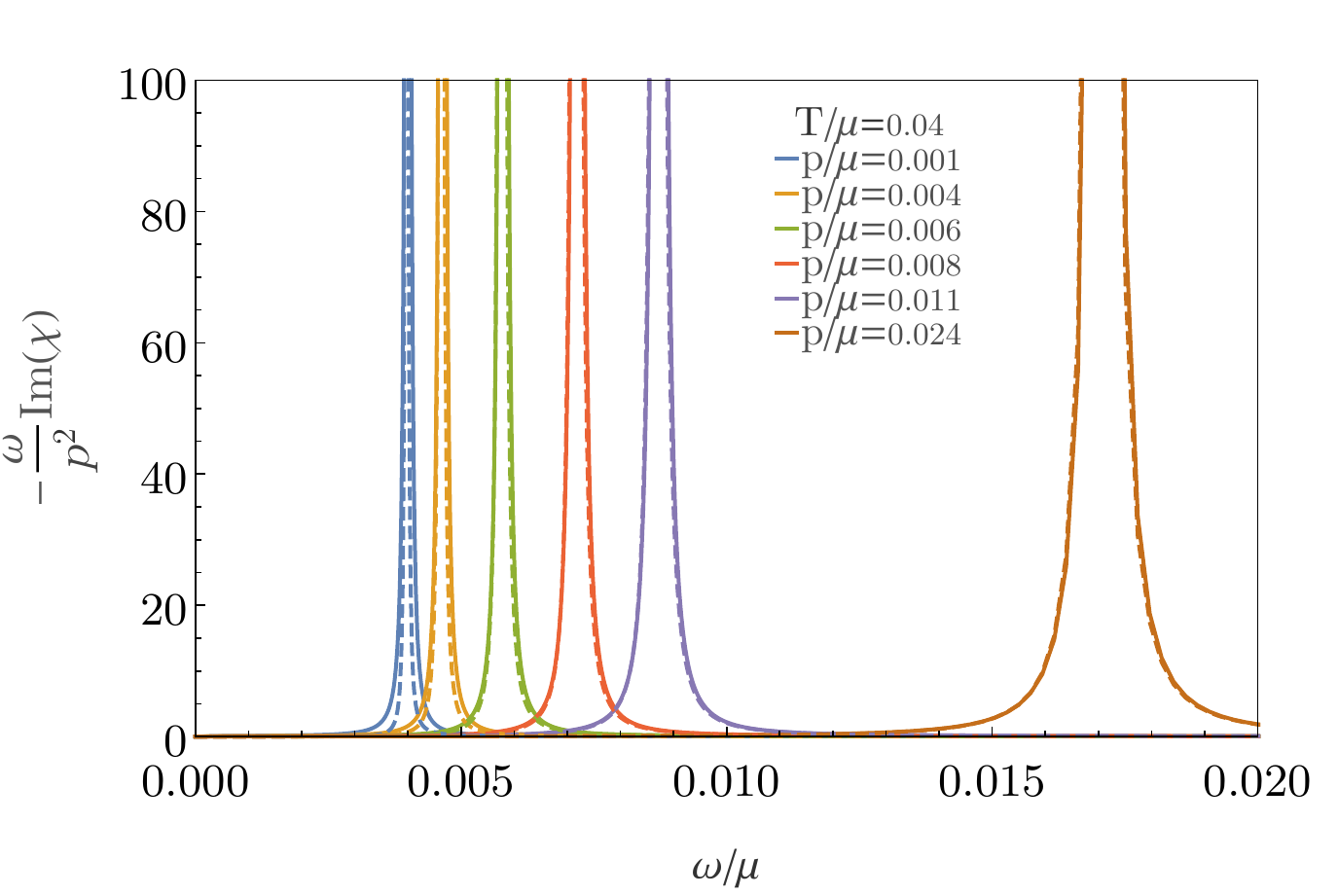}}
	\end{subfigure}
\vspace{-3mm}
\caption{\textbf{Density response dressed with the boundary Coulomb interaction}. Continuous  lines: translational symmetry is broken pseudo-spontaneously $\lambda/\mu=-10^{-4}$, $\alpha/\mu=0.1$. Dashed lines: translationally invariant system $\lambda/\mu=\alpha/\mu=0$. The effect of the Coulomb interaction is to gap the response at low momentum. The dressed response still displays an asymmetric peak which quickly becomes sound-dominated as momentum increases. }\label{fig:Dresed_plasmon}
\end{figure}

The shift and gapping of the hydrodynamic modes of the density correlator towards the real line and with larger real part suggests the dressed density response at vanishing momentum displays a gapped excitation which is narrower compared to the neutral response. Indeed, this is the behaviour seen by comparing the continuous lines of  Figs. \ref{fig:Dresed_plasmon} and \ref{fig:density_PseudoSpontaneous}. In Fig. \ref{fig:Dresed_plasmon}, the dashed lines correspond the response of the equivalent translationally invariant system at the same temperature and momentum. As the momentum or temperature increases, the effects of translational symmetry breaking in the dressed response  are again negligible.

\section{Conclusions}
The surprisingly over-damped plasmons observed in strange metals have revitalized the research on the study of the density response in these systems \cite{Nucker1989,Nucker1991,Fink1994,Mitrano2017,Husain2019}. As usual in this context, measurements in strange metals challenge the intuition based on the Fermi-liquid paradigm and open up the possibility to new exciting mechanisms \cite{Zaanen:2018}. Holography provides a powerful tool where one can hope to identify universal qualitative features caused by strong interactions.   Here, we have extended previous efforts on describing the phenomenology of holographic plasmons \cite{Aronsson2017,Aronsson2018,Gran2018,Gran2018b,Mauri2018, Romero-Bermudez2018}. In particular, we have carried the first study of the finite-momentum density response in a holographic setup in which translational symmetry is broken pseudo-spontaneously. The model we chose is metallic at low temperature, has infinite dynamical Lifshitz and Hyperscaling exponents, and displays linear scaling of entropy at low temperature. For some range of temperature, the zero-momentum optical conductivity has a pinned peak at finite frequency, which has been suggested to be behind the spectral shift to finite frequencies in many strange metals when they become  bad metals  at  large temperature \cite{Delacretaz2016}. However, contrary to experimental evidence, the location of the pinned peak in the conductivity in our model has a  non-monotonic dependence on temperature.  We also study the effect of temperature and momentum on the density-density  correlation function. When translations are broken, the neutral and dressed density responses  at low momentum are not dominated by the sound-mode. Instead, a broad but distinct asymmetric peak is observed in the density responses. We have also observed that the attenuation of this peak depends non-monotonically on momentum and temperature, similarly to the non-monotonic temperature-dependence of the location of the pinned peak in optical the conductivity. 

A final note regarding the connection to the over-damped plasmon measured experimentally is in order  \cite{Nucker1989,Nucker1991,Fink1994,Mitrano2017,Husain2019}. Previous measurements of the optical conductivity in materials  where the plasmon is being measured, like the  Bi-2212 and Bi-2201 families, indicate the existence of  spectral weight transfer to finite frequencies \cite{Lupi2000,Hwang2007}. This connects partially to the phenomenology of the model used here, where a peak at finite frequency is also seen.  However, the location of the pinned peak in our model depends non-monotonically on temperature. On the other hand, the spectral weight shift observed in the experiments occurs gradually as a function of temperature when entering the bad-metallic phase.  Therefore, this puts into question whether the momentum and temperature attenuation inversion of the holographic plasmons described here is relevant for the density response measured in these materials. Moreover, the existence of a non-Drude-like optical conductivity in these materials is currently been revisited by various experimental groups. In order to connect with experiments, it would be interesting to find holographic models where the AC-conductivity is Drude-like, but the dressed density response does not display big and narrow peaks \cite{Romero-Bermudez2019}. It would also be interesting to explore holographically the mismatch between the longitudinal and transverse density responses in the low momentum regime observed in cuprates \cite{Setty2018}.
\acknowledgments
It is a pleasure to thank Daniel Are\'an, Andrea Amoretti, 	Pau Figueras, Erik van Heumen, Alexander Krikun, Daniele Musso,  Koenraad Schalm, Jan Zaanen  and Vaios Ziogas for interesting discussions on various theoretical and practical aspects of this research. 
This work was supported by the Netherlands Organization for Scientific Research/Ministry of Science and Education (NWO/OCW), by the Foundation for Research into Fundamental Matter (FOM).

\appendix
\section{Equations of motion and boundary conditions}
\label{app:EOMs}

The equations of motion for the background fields following from the action \eqref{eq:action} are:
	\begin{align}
	&R_{\mu\nu}+3 g_{\mu\nu} = {Z(\phi)\over2} \left(F_{\mu\rho}F_{\nu}^{\phantom{\nu}\rho}-{F^2\over4}g_{\mu\nu}\right)+\half \p_\mu\phi\p_\nu\phi+V(\phi)+{Y(\phi)\over2}\sum_{I=1}^2\p_\mu\psi_I\p_\nu\psi_I,\notag\\
	&\cov_\mu (Z(\phi)F^{\mu\nu})= 0\,,\\
	 & \cov^2\phi {-}V'(\phi){-}{Z'(\phi)\over 4}F^2-{Y(\phi)\over2}\sum_{I=1}^2\p_\mu\psi_I\p^\mu\psi_I=0\,,\notag\\
	 & \cov_\mu(Y(\phi) \cov^\mu \psi_I)=0\,.\notag\\
	\end{align}
These equations result in a system of coupled ordinary equaitons which may be solved using the shooting method in the harmonic gauge \cite{Garcia-Garcia2016,Amoretti2018}. Here we choose to use the DeTruck method to solve the equations as a boundary value problem \cite{Headrick2010,Adam2012,Rangamani:2015, Andrade:2017b,Krikun2018}. We choose the Reissner-Nordstr\"om black hole as reference metric. Therefore, the temperature is set by the parameter $\bar{\mu}$ 
\begin{equation}\label{eq:T}
\frac{T}{\mu} = \frac{12 - \bar \mu^2}{16 \pi \mu},
\end{equation} 
and we can set $\bar{\mu} = \mu$ without loss of generality. We use the Newton's method to solve the resulting nonlinear elliptic equation in an unifrom grid. The boundary conditions are such that the metric is asymptotically AdS for $u\to0$: 
\begin{align}
Q_{tt}(u\to0)&\simeq 1 +\OO(u^2)\notag\\
Q_{uu}(u\to0)&\simeq 1 +\OO(u^2)\notag\\
Q_{xx}(u\to0)&\simeq 1 +\OO(u^2)\notag\\
A_t(u\to0)&\simeq \mu +\OO(u)\notag\\
\phi(u\to0)&\simeq \lambda +\OO(u)\notag\,,
\end{align}
where we choose $\lambda=0$ for to have spontaneously translations or $\lambda/\mu\ll1$  for pseudo-spontaneous breaking. In the horizon, we impose $Q_{tt}(u=1)=Q_{zz}(u=1)$ to have a static background, and the boundary conditions are the usual constraints on the expansion coefficients  of each field, obtained from expanding the equations of motion near the horizon.

\subsection{Linear response equations}\label{app:perturbations}

In order to evaluate the density-density correlator we first introduce the linear perturbations of all the fields with finite frequency and momentum:
\begin{equation}
\delta \varphi \to  e^{- i \omega t + i p x} \delta \varphi\,,\notag
\end{equation} 
and focus on the longitudinal modes
\begin{align}
\left.
\begin{cases}
\delta g_{tt}, \delta g_{tx}, \delta g_{tz}, \delta g_{xx}, \delta g_{xz}, \delta g_{yy}, \delta g_{zz},\delta A_t, \delta A_x, \delta A_z, \delta \phi, \delta \psi_x \notag
\end{cases}\hspace{-3mm}\right\}\,.
\end{align}
We impose the DeDonder gauge in $\delta g_{\mu \nu}$ and Lorentz gauge in $\delta A_\mu$ \cite{Rangamani:2015}:
\begin{align}
\cov^\mu\left(\delta g_{\mu\nu}-{\delta g^\alpha_{\phantom{\alpha}\alpha} \over 2} g^{(0)}_{\mu\nu} \right)=0\,, \ \cov^\mu \delta A_\mu=0\,,
\end{align}
where $g^{(0)}_{\mu\nu}$ is the background metric. This procedure results in 12 second order elliptic equations of motion.

In order to study the spontaneous symmetry breaking we redefine the fields
\begin{align}
\{ \delta  g_{tt}, \delta  g_{xx},  g_{yy} \}&= 
{(1-u)^{\sigma}\over u^2}\  \{ \delta \tilde g_{tt}, \delta \tilde g_{xx}, \delta \tilde g_{yy} \} \\
\delta  g_{uu}& = {(1-u)^{\sigma-2}\over u}\delta \tilde g_{uu}\\
\{\delta g_{tu},\delta g_{xu},\delta A_u\} & = {(1-u)^{\sigma-1}\over u}\{\delta \tilde g_{tu},\delta \tilde g_{xu},u\delta \tilde A_u\}\\
\{\delta A_t,\delta A_x, \delta g_{tx},\delta \psi_x, \phi \}&=(1-u)^{\sigma}\{\delta \tilde A_t,\delta \tilde A_x, \delta\tilde  g_{tx},u\tilde \phi\}\\
\delta  \psi_x &={(1-u)^{\sigma}\over u}\delta \tilde \psi_x\,,\label{eq:psi_BC}
\end{align}
where $\sigma=-i\omega/4\pi T$. The factor $(1-u)^\sigma$ imposes  ingoing boundary conditions are imposed in the horizon. Moreover, the other factors are chosen so that the first term in the expansion of the tilde-fields expand is a constant both in the UV and horizon. Dependending on whether we study the electric conductivity or the density response, we turn on a source $\delta \tilde A_x(u=0)=1$ or $\delta \tilde A_t(u=0)=1$, respectively. No other source is turned on (trivial Dirichlet boundary conditions at $u=0$). On the horizon, the boundary conditions are imposed as usual: we use expansion of the equations near $u=1$, which imposes constraints betweeen the expansion coefficients of the fields. These constraints result in mixed boundary conditions at $u=1$.
In order to study pseudo-spontaneous and explicit breaking of translational symmetry, the only change is the redefinition of Eq. \eqref{eq:psi_BC}, which now should be $\delta  \psi_x ={(1-z)^{\gamma}}\delta \tilde \psi_x$.

The equations are solved by linearizing and discretizing them which allows to recast the equations as a linear algebraic problem
\begin{equation}
\mathcal{M} \vec{f} = \vec{\cal{A}},
\end{equation}
where the right hand side corresponds to the sources being turned on. We also can obtain the  quasinormal modes as the solutions to the Sturm-Liuville problem 
\begin{equation}
\mathcal{M} \vec{f} = 0,
\end{equation} 
as the eigenvalues of $\mathcal{M}(\omega)$.

As explained in the main text, the plasmon quasinormal modes are obtained in a similar way, but using the mixed Robin boundary conditions \eqref{eq:Robin} \cite{Romero-Bermudez2018}. This results in a different matrix associated to the linear system of equations $\mathcal{M}_{V_\bp}(\omega)$.

\bibliographystyle{JHEP}
\bibliography{library}

\providecommand{\href}[2]{#2}\begingroup\raggedright\begin{thebibliography}{10}

\bibitem{Nozieres1999}
P.~Nozieres and D.~Pines, \emph{Theory Of Quantum Liquids}, Advanced Books
  Classics. Avalon Publishing, 1999.

\bibitem{Bozovic1987}
I.~Bozovic, D.~Kirillov, A.~Kapitulnik, K.~Char, M.~R. Hahn, M.~R. Beasley
  et~al., \emph{Optical measurements on oriented thin
  ${\mathrm{yba}}_{2}$${\mathrm{cu}}_{3}$${\mathrm{o}}_{7\mathrm{\ensuremath{-}}\mathrm{\ensuremath{\delta}}}$
  films: Lack of evidence for excitonic superconductivity},
  \href{https://doi.org/10.1103/PhysRevLett.59.2219}{\emph{Phys. Rev. Lett.}
  {\bfseries 59} (1987) 2219}.

\bibitem{Slakey1991}
F.~Slakey, M.~V. Klein, J.~P. Rice and D.~M. Ginsberg, \emph{Raman
  investigation of the
  ${\mathrm{yba}}_{2}$${\mathrm{cu}}_{3}$${\mathrm{o}}_{7}$ imaginary response
  function}, \href{https://doi.org/10.1103/PhysRevB.43.3764}{\emph{Phys. Rev.
  B} {\bfseries 43} (1991) 3764}.

\bibitem{Nucker1989}
N.~N{\"u}cker, H.~Romberg, S.~Nakai, B.~Scheerer, J.~Fink, Y.~F. Yan et~al.,
  \emph{Plasmons and interband transitions in bi 2 sr 2 ca cu 2 o 8},
  {\emph{Physical Review B} {\bfseries 39} (1989) 12379}.

\bibitem{Nucker1991}
N.~N{\"u}cker, U.~Eckern, J.~Fink and P.~M{\"u}ller, \emph{Long-wavelength
  collective excitations of charge carriers in high-t c superconductors},
  {\emph{Physical Review B} {\bfseries 44} (1991) 7155}.

\bibitem{Fink1994}
M.~{Knupfer}, G.~{Roth}, J.~{Fink}, J.~{Karpinski} and E.~{Kaldis},
  \emph{{Plasmon dispersion and the dielectric function in YBa $_{2}$Cu $_{4}$O
  $_{8}$ single crystals}},
  \href{https://doi.org/10.1016/0921-4534(94)90453-7}{\emph{Physica C
  Superconductivity} {\bfseries 230} (1994) 121}.

\bibitem{Mitrano2017}
M.~Mitrano, A.~A. Husain, S.~Vig, A.~Kogar, M.~S. Rak, S.~I. Rubeck et~al.,
  \emph{{Anomalous density fluctuations in a strange metal}},
  \href{https://doi.org/10.1073/pnas.1721495115}{\emph{Proc. Natl. Acad. Sci.}
  {\bfseries 115} (2018) 5392}
  [\href{https://arxiv.org/abs/1708.01929}{{\ttfamily 1708.01929}}].

\bibitem{Husain2019}
A.~Husain, M.~Mitrano, M.~S. Rak, S.~Rubeck, B.~Uchoa, J.~Schneeloch et~al.,
  \emph{{Crossover of Charge Fluctuations across the Strange Metal Phase
  Diagram}},  \href{https://arxiv.org/abs/1903.04038v1}{{\ttfamily
  1903.04038v1}}.

\bibitem{Amoretti2018}
A.~Amoretti, D.~Are{\'{a}}n, B.~Gout{\'{e}}raux and D.~Musso, \emph{{A
  holographic strange metal with slowly fluctuating translational order}},
  \href{https://arxiv.org/abs/1812.08118}{{\ttfamily 1812.08118}}.

\bibitem{Son2005}
D.~T. Son, \emph{{Effective Lagrangian and Topological Interactions in
  Supersolids}},
  \href{https://doi.org/10.1103/PhysRevLett.94.175301}{\emph{Phys. Rev. Lett.}
  {\bfseries 94} (2005) 175301}
  [\href{https://arxiv.org/abs/0501658v2}{{\ttfamily 0501658v2}}].

\bibitem{Nicolis2013}
A.~Nicolis, R.~Penco and R.~A. Rosen, \emph{{Relativistic fluids, superfluids,
  solids, and supersolids from a coset construction}},
  \href{https://doi.org/10.1103/PhysRevD.89.045002}{\emph{Phys. Rev. D}
  {\bfseries 89} (2014) 045002}
  [\href{https://arxiv.org/abs/1307.0517}{{\ttfamily 1307.0517}}].

\bibitem{Baggioli2014}
M.~Baggioli and O.~Pujol{\`{a}}s, \emph{{Electron-Phonon Interactions,
  Metal-Insulator Transitions, and Holographic Massive Gravity}},
  \href{https://doi.org/10.1103/PhysRevLett.114.251602}{\emph{Phys. Rev. Lett.}
  {\bfseries 114} (2015) 251602}
  [\href{https://arxiv.org/abs/1411.1003}{{\ttfamily 1411.1003}}].

\bibitem{Andrade2017}
T.~Andrade, M.~Baggioli, A.~Krikun and N.~Poovuttikul, \emph{{Pinning of
  longitudinal phonons in holographic helical crystals}},
  \href{https://arxiv.org/abs/1708.08306}{{\ttfamily 1708.08306}}.

\bibitem{Amoretti2017}
A.~Amoretti, D.~Are{\'{a}}n, B.~Gout{\'{e}}raux and D.~Musso, \emph{{Effective
  holographic theory of charge density waves}},
  \href{https://doi.org/10.1103/PhysRevD.97.086017}{\emph{Phys. Rev. D}
  {\bfseries 97} (2018) 086017}
  [\href{https://arxiv.org/abs/1711.06610}{{\ttfamily 1711.06610}}].

\bibitem{Amoretti2017a}
A.~Amoretti, D.~Are{\'{a}}n, R.~Argurio, D.~Musso and L.~A.~P. Zayas, \emph{{A
  holographic perspective on phonons and pseudo-phonons}},
  \href{https://doi.org/10.1007/JHEP05(2017)051}{\emph{J. High Energy Phys.}
  {\bfseries 2017} (2017) 51}
  [\href{https://arxiv.org/abs/1611.09344}{{\ttfamily 1611.09344}}].

\bibitem{Alberte2018}
L.~Alberte, M.~Ammon, M.~Baggioli, A.~Jim{\'{e}}nez and O.~Pujol{\`{a}}s,
  \emph{{Black hole elasticity and gapped transverse phonons in holography}},
  \href{https://doi.org/10.1007/JHEP01(2018)129}{\emph{J. High Energy Phys.}
  {\bfseries 2018} (2018) 129}
  [\href{https://arxiv.org/abs/1708.08477}{{\ttfamily 1708.08477}}].

\bibitem{Alberte2018a}
L.~Alberte, M.~Ammon, A.~Jim{\'{e}}nez-Alba, M.~Baggioli and O.~Pujol{\`{a}}s,
  \emph{{Holographic Phonons}},
  \href{https://doi.org/10.1103/PhysRevLett.120.171602}{\emph{Phys. Rev. Lett.}
  {\bfseries 120} (2018) 171602}
  [\href{https://arxiv.org/abs/1711.03100}{{\ttfamily 1711.03100}}].

\bibitem{Musso2018}
D.~Musso, \emph{{Simplest phonons and pseudo-phonons in field theory}},
  \href{https://arxiv.org/abs/1810.01799}{{\ttfamily 1810.01799}}.

\bibitem{Andrade2018}
T.~Andrade and A.~Krikun, \emph{{Coherent vs incoherent transport in
  holographic strange insulators}},
  \href{https://arxiv.org/abs/1812.08132}{{\ttfamily 1812.08132}}.

\bibitem{Donos2019}
A.~Donos and C.~Pantelidou, \emph{{Holographic transport and density waves}},
  \href{https://arxiv.org/abs/1903.05114}{{\ttfamily 1903.05114}}.

\bibitem{Karch:2009}
A.~Karch, D.~T. Son and A.~O. Starinets, \emph{{Holographic Quantum Liquid}},
  \href{https://doi.org/10.1103/PhysRevLett.102.051602}{\emph{Phys. Rev. Lett.}
  {\bfseries 102} (2009) 051602}
  [\href{https://arxiv.org/abs/0806.3796}{{\ttfamily 0806.3796}}].

\bibitem{Kulaxizi2008}
M.~{Kulaxizi} and A.~{Parnachev}, \emph{{Remarks on Fermi liquid from
  holography}}, \href{https://doi.org/10.1103/PhysRevD.78.086004}{\emph{Phys.
  Rev. D} {\bfseries 78} (2008) 086004}
  [\href{https://arxiv.org/abs/0808.3953}{{\ttfamily 0808.3953}}].

\bibitem{Kulaxizi2009}
M.~{Kulaxizi} and A.~{Parnachev}, \emph{{Holographic responses of fermion
  matter}},
  \href{https://doi.org/10.1016/j.nuclphysb.2009.02.016}{\emph{Nuclear Physics
  B} {\bfseries 815} (2009) 125}
  [\href{https://arxiv.org/abs/0811.2262}{{\ttfamily 0811.2262}}].

\bibitem{Hoyos2010}
C.~Hoyos, A.~O'Bannon and J.~M.~S. Wu, \emph{{Zero sound in strange metallic
  holography}}, \href{https://doi.org/10.1007/JHEP09(2010)086}{\emph{J. High
  Energy Phys.} {\bfseries 9} (2010) 86}
  [\href{https://arxiv.org/abs/1007.0590}{{\ttfamily 1007.0590}}].

\bibitem{Kaminski2010}
M.~{Kaminski}, K.~{Landsteiner}, J.~{Mas}, J.~P. {Shock} and
  J.~{Tarr{\'{\i}}o}, \emph{{Holographic operator mixing and quasinormal modes
  on the brane}}, \href{https://doi.org/10.1007/JHEP02(2010)021}{\emph{J. High
  Energy Phys.} {\bfseries 2} (2010) 21}
  [\href{https://arxiv.org/abs/0911.3610}{{\ttfamily 0911.3610}}].

\bibitem{Davison:2011}
R.~A. Davison and A.~O. Starinets, \emph{{Holographic zero sound at finite
  temperature}}, \href{https://doi.org/10.1103/PhysRevD.85.026004}{\emph{Phys.
  Rev. D} {\bfseries 85} (2012) 026004}
  [\href{https://arxiv.org/abs/1109.6343}{{\ttfamily 1109.6343}}].

\bibitem{Gushterov2018}
N.~I. Gushterov, A.~O'Bannon and R.~Rodgers, \emph{{Holographic zero sound from
  spacetime-filling branes}},
  \href{https://doi.org/10.1007/JHEP10(2018)076}{\emph{J. High Energy Phys.}
  {\bfseries 10} (2018) 76}.

\bibitem{Edalati2010c}
M.~Edalati, J.~I. Jottar and R.~G. Leigh, \emph{{Holography and the sound of
  criticality}}, \href{https://doi.org/10.1007/JHEP10(2010)058}{\emph{J. High
  Energy Phys.} {\bfseries 10} (2010) 58}
  [\href{https://arxiv.org/abs/arXiv:1005.4075v2}{{\ttfamily
  arXiv:1005.4075v2}}].

\bibitem{Davison2011b}
R.~A. Davison and N.~K. Kaplis, \emph{{Bosonic excitations of the AdS 4
  Reissner-Nordstrom black hole}},
  \href{https://doi.org/10.1007/JHEP12(2011)037}{\emph{J. High Energy Phys.}
  {\bfseries 12} (2011) 37} [\href{https://arxiv.org/abs/1111.0660}{{\ttfamily
  1111.0660}}].

\bibitem{Aronsson2017}
M.~Aronsson, U.~Gran and T.~Zingg, \emph{Holographic plasmons},
  \href{https://arxiv.org/abs/1712.05672}{{\ttfamily 1712.05672}}.

\bibitem{Aronsson2018}
M.~Aronsson, U.~Gran and T.~Zingg, \emph{Plasmons in holographic graphene},
  \href{https://arxiv.org/abs/1804.02284}{{\ttfamily 1804.02284}}.

\bibitem{Gran2018}
U.~Gran, M.~Torns{\"o} and T.~Zingg, \emph{Exotic holographic dispersion},
  \href{https://arxiv.org/abs/1808.05867}{{\ttfamily 1808.05867}}.

\bibitem{Gran2018b}
U.~Gran, M.~Tornso and T.~Zingg, \emph{{Holographic Response of Electron
  Clouds}},  \href{https://arxiv.org/abs/1810.11416}{{\ttfamily 1810.11416}}.

\bibitem{Mauri2018}
E.~Mauri and H.~Stoof, \emph{{Screening of Coulomb interactions in
  Holography}},  \href{https://arxiv.org/abs/1811.11795}{{\ttfamily
  1811.11795}}.

\bibitem{Romero-Bermudez2018}
A.~Krikun, A.~Romero-Berm{\'{u}}dez, K.~Schalm and J.~Zaanen, \emph{{Anomalous
  attenuation of plasmons in strange metals and holography}},
  \href{https://doi.org/10.1103/PhysRevB.99.235149}{\emph{Phys. Rev. B -
  Editor's Choice} {\bfseries 99} (2019) 235149}
  [\href{https://arxiv.org/abs/1812.03968}{{\ttfamily 1812.03968}}].

\bibitem{Fradkin2015}
E.~Fradkin, S.~A. Kivelson and J.~M. Tranquada, \emph{{Colloquium: Theory of
  intertwined orders in high temperature superconductors}},
  \href{https://doi.org/10.1103/RevModPhys.87.457}{\emph{Rev. Mod. Phys.}
  {\bfseries 87} (2015) 457}
  [\href{https://arxiv.org/abs/arXiv:1407.4480v3}{{\ttfamily
  arXiv:1407.4480v3}}].

\bibitem{Lubensky_book}
P.~M. Chaikin and T.~C. Lubensky, \emph{Principles of Condensed Matter
  Physics}. Cambridge University Press, 1995,
  \href{https://doi.org/10.1017/CBO9780511813467}{10.1017/CBO9780511813467}.

\bibitem{Delacretaz2016}
L.~V. Delacr{\'{e}}taz, B.~Gout{\'{e}}raux, S.~A. Hartnoll and A.~Karlsson,
  \emph{{Bad Metals from Density Waves}},
  \href{https://arxiv.org/abs/1612.04381}{{\ttfamily 1612.04381}}.

\bibitem{Delacretaz2017a}
L.~V. Delacr{\'{e}}taz, B.~Gout{\'{e}}raux, S.~A. Hartnoll and A.~Karlsson,
  \emph{{Theory of hydrodynamic transport in fluctuating electronic charge
  density wave states}},
  \href{https://doi.org/10.1103/PhysRevB.96.195128}{\emph{Phys. Rev. B}
  {\bfseries 96} (2017) 195128}
  [\href{https://arxiv.org/abs/1702.05104}{{\ttfamily 1702.05104}}].

\bibitem{Keimer2014}
B.~Keimer, S.~A. Kivelson, M.~R. Norman, S.~Uchida and J.~Zaanen, \emph{{From
  quantum matter to high-temperature superconductivity in copper oxides}},
  \href{https://doi.org/10.1038/nature14165}{\emph{Nature} {\bfseries 518}
  (2015) 179}.

\bibitem{ZaanenBook}
J.~Zaanen, Y.-W. Sun, Y.~Liu and K.~Schalm, \emph{Holographic Duality in
  Condensed Matter Physics}. Cambridge Univ. Press, 2015.

\bibitem{Hartnoll:2016Rev}
S.~A. Hartnoll, A.~Lucas and S.~Sachdev, \emph{{Holographic quantum matter}},
  \href{https://arxiv.org/abs/1612.07324}{{\ttfamily 1612.07324}}.

\bibitem{Zaanen:2018}
J.~Zaanen, \emph{{Planckian dissipation, minimal viscosity and the transport in
  cuprate strange metals}},  \href{https://arxiv.org/abs/1807.10951}{{\ttfamily
  1807.10951}}.

\bibitem{Hussey2004}
N.~E. {Hussey ‖}, K.~Takenaka and H.~Takagi, \emph{{Universality of the
  Mott--Ioffe--Regel limit in metals}},
  \href{https://doi.org/10.1080/14786430410001716944}{\emph{Philos. Mag.}
  {\bfseries 84} (2004) 2847}.

\bibitem{Lupi2000}
S.~Lupi, P.~Calvani, M.~Capizzi and P.~Roy, \emph{{Evidence of two species of
  carriers from the far-infrared reflectivity of Bi2Sr2CuO6}},
  \href{https://doi.org/10.1103/PhysRevB.62.12418}{\emph{Phys. Rev. B}
  {\bfseries 62} (2000) 418}.

\bibitem{Hwang2007}
J.~Hwang, T.~Timusk and G.~D. Gu, \emph{{Doping dependent optical properties of
  Bi2Sr 2CaCu2O8+$\delta$}},
  \href{https://doi.org/10.1088/0953-8984/19/12/125208}{\emph{J. Phys. Condens.
  Matter} {\bfseries 19} (2007) }.

\bibitem{Donos2014b}
A.~Donos and J.~P. Gauntlett, \emph{{Holographic Q-lattices}},
  \href{https://doi.org/10.1007/JHEP04(2014)040}{\emph{J. High Energy Phys.}
  {\bfseries 04} (2014) 40} [\href{https://arxiv.org/abs/1311.3292}{{\ttfamily
  1311.3292}}].

\bibitem{Donos2014e}
A.~Donos and J.~P. Gauntlett, \emph{{Novel metals and insulators from
  holography}}, \href{https://doi.org/10.1007/JHEP06(2014)007}{\emph{J. High
  Energy Phys.} {\bfseries 2014} (2014) }
  [\href{https://arxiv.org/abs/1401.5077}{{\ttfamily 1401.5077}}].

\bibitem{Garcia-Garcia2016a}
A.~M. Garc{\'{i}}a-Garc{\'{i}}a and A.~Romero-Berm{\'{u}}dez, \emph{{Drude
  weight and Mazur-Suzuki bounds in holography}},
  \href{https://doi.org/10.1103/PhysRevD.93.066015}{\emph{Phys. Rev. D}
  {\bfseries 93} (2016) } [\href{https://arxiv.org/abs/1512.04401}{{\ttfamily
  1512.04401}}].

\bibitem{Donos2018}
A.~Donos, J.~P. Gauntlett, T.~Griffin and V.~Ziogas, \emph{{Incoherent
  transport for phases that spontaneously break translations}},
  \href{https://doi.org/10.1007/JHEP04(2018)053}{\emph{J. High Energy Phys.}
  {\bfseries 2018} (2018) }
  [\href{https://arxiv.org/abs/arXiv:1801.09084v3}{{\ttfamily
  arXiv:1801.09084v3}}].

\bibitem{Romero-Bermudez2019}
T.~Andrade, A.~Krikun and A.~Romero-Berm{\'{u}}dez, \emph{{To appear:
  Incoherent density response in holographic insulators}}.

\bibitem{Setty2018}
C.~Setty, B.~Padhi, K.~Limtragool, P.~Abbamonte, A.~A. Husain, M.~Mitrano
  et~al., \emph{{Inequivalence of the zero-momentum Limits of Transverse and
  Longitudinal Dielectric Response in the Cuprates}},
  \href{https://arxiv.org/abs/1803.05439}{{\ttfamily 1803.05439}}.

\bibitem{Garcia-Garcia2016}
A.~M. Garc{\'{i}}a-Garc{\'{i}}a, B.~Loureiro and A.~Romero-Berm{\'{u}}dez,
  \emph{{Transport in a gravity dual with a varying gravitational coupling
  constant}}, \href{https://doi.org/10.1103/PhysRevD.94.086007}{\emph{Phys.
  Rev. D} {\bfseries 94} (2016) 086007}
  [\href{https://arxiv.org/abs/1606.01142}{{\ttfamily 1606.01142}}].

\bibitem{Headrick2010}
M.~Headrick, S.~Kitchen and T.~Wiseman, \emph{A new approach to static
  numerical relativity and its application to kaluza--klein black holes},
  {\emph{Classical and Quantum Gravity} {\bfseries 27} (2010) 035002}
  [\href{https://arxiv.org/abs/0905.1822}{{\ttfamily 0905.1822}}].

\bibitem{Adam2012}
A.~Adam, S.~Kitchen and T.~Wiseman, \emph{A numerical approach to finding
  general stationary vacuum black holes}, {\emph{Classical and Quantum Gravity}
  {\bfseries 29} (2012) 165002}
  [\href{https://arxiv.org/abs/1105.6347}{{\ttfamily 1105.6347}}].

\bibitem{Rangamani:2015}
M.~Rangamani, M.~Rozali and D.~Smyth, \emph{{Spatial Modulation and
  Conductivities in Effective Holographic Theories}},
  \href{https://doi.org/10.1007/JHEP07(2015)024}{\emph{J. High Energy Phys.}
  {\bfseries 07} (2015) 024}
  [\href{https://arxiv.org/abs/1505.05171}{{\ttfamily 1505.05171}}].

\bibitem{Andrade:2017b}
T.~Andrade, A.~Krikun, K.~Schalm and J.~Zaanen, \emph{{Doping the holographic
  Mott insulator}},
  \href{https://doi.org/10.1038/s41567-018-0217-6}{\emph{Nature Physics}
  {\bfseries 14} (2018) 1049}
  [\href{https://arxiv.org/abs/1710.05791}{{\ttfamily 1710.05791}}].

\bibitem{Krikun2018}
A.~Krikun, \emph{Numerical solution of the boundary value problems for partial
  differential equations. crash course for holographer},
  \href{https://arxiv.org/abs/1801.01483}{{\ttfamily 1801.01483}}.

\end{thebibliography}\endgroup

\end{document}